\documentclass[11pt]{article}

\usepackage[utf8]{inputenc}
\usepackage[T1]{fontenc}
\usepackage{lmodern}
\usepackage[margin=1in]{geometry}
\usepackage{amsmath,amsthm,amsfonts,amssymb}
\usepackage{graphicx}
\usepackage{float}
\usepackage{enumitem}
\usepackage{siunitx}
\usepackage{booktabs}
\usepackage{etoolbox}
\AtBeginEnvironment{table}{\small\setlength{\tabcolsep}{4pt}}
\usepackage[numbers,sort&compress]{natbib}
\usepackage[colorlinks,citecolor=blue,urlcolor=blue,linkcolor=blue]{hyperref}

\providecommand{\tabularnewline}{\\}
\newcommand{\mc}[1]{\multicolumn{1}{c}{#1}}
\newcommand{\bb}{\boldsymbol{\beta}}
\newcommand{\bL}{{\Lambda}}
\newcommand{\bl}{\boldsymbol{\lambda}}
\newcommand{\bt}{\boldsymbol{\theta}}
\newcommand{\bO}{\boldsymbol{\Omega}}
\newcommand{\bS}{\boldsymbol{\Sigma}}
\newcommand{\bV}{\mathbf{V}}
\newcommand{\bX}{\mathbf{X}}
\newcommand{\bx}{\mathbf{x}}

\theoremstyle{plain}

\newtheorem{prop}{Proposition}
\newtheorem{lem}{Lemma}

\newtheorem{theorem}{Theorem}[section]

\theoremstyle{remark}

\title{A Robust Approach for Regression Analysis of Panel Count Data with Time-varying Covariates}
\author{
Dayu Sun\thanks{Department of Biostatistics and Bioinformatics, Emory University, GA 30322, U.S.A. Email: \texttt{dayusun@iu.edu}}
\and
Yuanyuan Guo\thanks{Department of Biostatistics and Bioinformatics, Duke University, NC 27710, U.S.A. Email: \texttt{yuanyuan.guo@duke.edu}}
\and
Yang Li\thanks{Department of Biostatistics and Health Data Science, Indiana University, Indianapolis, IN 46202, U.S.A. Email: \texttt{yli18@iu.edu}}
\and
Wanzhu Tu\thanks{Department of Biostatistics and Health Data Science, Indiana University, Indianapolis, IN 46202, U.S.A. Email: \texttt{wtu1@iu.edu}}
\and
Jianguo Sun\thanks{Corresponding author. Department of Statistics, University of Missouri, MO 65211, U.S.A. Email: \texttt{sunj@missouri.edu}}
}
\date{}

\begin{document}
\maketitle

\begin{abstract}
The validity of statistical inference for panel count data with time-varying covariates depends on the correct specification of within-subject correlation structures; misspecification often leads to questionable inference. To alleviate, robust inference has been proposed for mean models, which implicitly assume monotone mean functions. When covariate values fluctuate with time, however, the assumed monotonicity becomes unrealistic. In this research, we propose a robust inference based on rate models that are free of such constraints. Since the asymptotic variance has no closed form under the rate model, we further develop computationally efficient robust variance estimators using the Expectation-Maximization (EM) algorithm, thus sidestepping the need for computationally intensive numerical methods, which could undermine the robustness. Rigorous theoretical development is provided in support of parameter estimation and inference. Extensive simulation studies demonstrate the superiority of the proposed method. We present a real clinical application to illustrate the use of the proposed method.

\end{abstract}
\section{Introduction\label{sec:intro}}

In recurrent event studies, precise event times are typically unknown when observations are made periodically at discrete points. Instead, one only obtains counts of events that occur between the consecutive observation points. Such data are referred to as panel count data. Panel count data are abundant in event history studies \citep{Balakrishnan2008, Balakrishnan2009,Hu2009,Jiang2018,Sun2013, Zhao2018}. A common interest of panel count data analysis is to determine the effects of covariates on the event rate. In practice, covariates could vary with time, or there may exist an interaction between a time-invariant covariate and time, thus further complicating the analysis. An application that motivated the current research is the Young Women's Project (YWP), an epidemiological study of the sexual behaviors' effects on sexually transmitted infections (STIs) \citep{Ghosh2009,Tu2009}. The YWP yielded panel count data on STIs as the exact times of infection were unknown due to its quarterly testing schedule. The subjects' sexual activities since the last visit often varied with time and were summarized as time-varying covariates. 

Various models for panel count data with time-varying covariates have been proposed. Let $N\left(t\right)$ be the counting process for the total number of recurrent events up to time $t$; the process jumps with increment 1. For the analysis, a natural approach is to characterize the relationship between $N\left(t\right)$ and
a vector of covariates denoted by $\bX\left(t\right)\in\mathbb{R}^{p}$ in the following intensity process model 
\begin{equation}
	E\left[dN\left(t\right)|\mathcal{F}_{t-}\right]=\exp(\bb^{\top}\bX\left(t\right))d\bL\left(t\right), \label{eq:intensitymodel}
\end{equation}
\citep{Andersen1982}. In the above model, $dN\left(t\right)=N\left(t+dt-\right)-N\left(t-\right)$ denotes
the increment of $N$ within the small interval $[t,t+dt)$, $\mathcal{F}_{t}$ is the $\sigma$-field generated by $\{ N\left(s\right),\bX\left(s\right),\allowdisplaybreaks 0\le s\le t\}$, $\bb\in\mathbb{R}^{p}$ represents covariate effects, and $\bL (t)$ is an unknown baseline cumulative intensity function. Model \eqref{eq:intensitymodel} specifies the covariate effects to be proportional to the instantaneous probability of event occurrence at time $t$ given the history of $N\left(\cdot\right)$ and $\bX\left(\cdot\right)$ before $t$. Implicitly, model \eqref{eq:intensitymodel} requires a modulated Poisson process where the intensity at $t$ depends solely on $\bX\left(t\right)$ at time $t$ \citep{Cook2007a}. However, the Poisson assumption is often unrealistic because it implies the event counts of the same subject from non-overlapping intervals are mutually independent, which is generally untrue in longitudinal studies such as YWP. Thus, the validity of inference in panel count data analysis is contingent on the correct specification of the within-subject correlation structures.

To relax the Poisson assumption, two semiparametric models have been considered for panel count data regression.  One is the proportional rates model 
\begin{equation}
	E\left[dN\left(t\right)|\bX\left(t\right)\right]=\exp(\bb^{\top}\bX\left(t\right))d\bL\left(t\right), \label{eq:ratemodel}
\end{equation}
\citep{Lin2000}, and the other is the proportional means model 
\begin{equation}
	E\left[N\left(t\right)|\bX\left(t\right)\right]=\mu\left(t\right)\exp(\bb^{\top}\bX\left(t\right)), \label{eq:meanmodel}
\end{equation}
\citep{Sun2000,Hu2003}.
Here, $\bb$ is defined as in model \eqref{eq:intensitymodel}, $\bL\left(t\right)$ is an unspecified non-decreasing cumulative
baseline rate function, and $\mu\left(t\right)$ is an unspecified non-decreasing baseline mean function.  It is easy to see that model (\ref{eq:ratemodel}) is equivalent to model (\ref{eq:intensitymodel}) under the Poisson assumption, and models (\ref{eq:ratemodel}) and (\ref{eq:meanmodel}) are the same with $\mu\left(t\right)=\bL\left(t\right)$ when the covariates are time-invariant.  But with time-varying covariates, models (\ref{eq:ratemodel}) and (\ref{eq:meanmodel}) are generally not equivalent, and different inference procedures are therefore needed.

To our knowledge, the existing robust estimation procedures are all based on the means model (\ref{eq:meanmodel}). For example, \citet{Sun2000} and \citet{Hu2003} discussed the estimation for model (\ref{eq:meanmodel}), and they proposed to ascertain covariate effects by using estimating equation-based methods without relying on the Poisson assumption. Under the Poisson assumption, \citet{Wellner2007} investigated the full and pseudo maximum likelihood estimations and proved that the resulting estimators are still consistent when the Poisson assumption does not hold. \citet{Hua2014} later generalized Wellner and Zhang's approach by specifically modeling the overdispersion by adding a parametric Gamma-frailty and showed that the extended method is robust with respect to the parametric assumption. Similar methods were also discussed by \citet{Zhao2018} and \citet{Wang2019}. However, all of the above methods are for the situation of time-invariant covariates, thus are not readily applicable to studies such as the YWP.

For panel count data with time-varying covariates, robust methods are limited, and they are generally developed within the framework of proportional mean model \eqref{eq:meanmodel}  \citep{Deng2013,Li2010,Li2013}. A specific challenge for model \eqref{eq:meanmodel} is that with time-varying covariates, it is impossible to guarantee the monotonicity of the baseline mean function $\mu(t)$. In fact, most robust methods under model \eqref{eq:meanmodel} produce non-monotonic estimates of the baseline mean function, and thus severely undermining their practical utility. 

In contrast, the proportional rate model (\ref{eq:ratemodel}) is not subject to this limitation as it only requires the right-hand side of (\ref{eq:ratemodel}) to be positive, and the time-varying covariates can therefore be readily accommodated. Furthermore, model (\ref{eq:ratemodel}) has an appealing feature that $E\left[dN\left(t\right)|\bX\left(t\right)\right]$ stands for the marginal instantaneous probability of an event occurring at time $t$ given the covariates at $t$ or the 
instantaneous relative risk, analogous to the hazard function in survival analysis. 

The rate model (\ref{eq:ratemodel}), despite its inherent advantage, has not been carefully studied for panel count regression with time-varying covariates. \citet{Zeng2020} investigated the maximum likelihood estimation of model (\ref{eq:intensitymodel}) with time-varying covariates, which could apply to model (\ref{eq:ratemodel}) under the Poisson assumption. Their method, however, relies heavily on the Poisson assumption and the correct specification of within-subject correlations, which is difficult to justify in practical data analysis.  Additionally, the lack of a closed-form estimator for their asymptotic variance also presents a challenge: Without an explicit formulation for the asymptotic variances of $\hat{\bb}$, it will be difficult to carry out inference. Numerical derivatives of the profile log-likelihood function \citep{Murphy1997,Murphy2000} may provide some relief, and that approach has been used by \citet{Su2016}, \citet{Zeng2016} and \citet{Zeng2020} with slight modifications. Nevertheless, the profile likelihood method is not only computationally intensive but also vulnerable to model misspecification.  

To overcome these difficulties, we develop a robust likelihood-based estimation procedure for model (\ref{eq:ratemodel}) with time-varying covariates. Key contributions of the paper are: 

(i) We provide rigorous proofs showing that consistency and asymptotically normality of the robust estimator can be achieved without relying on correctly specified within-subject correlation structures. For implementation, we propose an efficient Expectation-Maximization (EM) algorithm.

(ii) We propose two novel variance estimators in tractable forms, thus greatly improving computational efficiency. One is robust to the within-subject correlation assumption, and the other serves as a more efficient substitute for the profile likelihood variance estimates. Notably, the robust variance estimator is a sandwich-type estimator based on the efficient score constructed under the working Poisson assumption in a fashion analogous to \citet{Lin2000} and \citet{Wellner2007}. But unlike the published work, the forms of these two variance estimators are based on the posterior score function of $\bb$ in the EM algorithm. Specifically, since the least-favorable direction has no explicit form, we estimate it directly by the derivative of the profiled baseline rate function.

(iii) We provide theoretical justifications for the variance estimators by establishing a general connection between the EM algorithm and the efficient score function of $\bb$ \emph{regardless of distribution assumptions}, based on the modern empirical process theory and semiparametric theory. To our knowledge, there is scarce literature on the consistency of variance estimators based on the EM algorithm under semiparametric settings. The existing literature, e.g. \citet{Oakes1999} and \citet{Louis1982}, only derived the population forms of some estimator variance based on the EM algorithm for parametric models. The proposed variance estimation methods could be generalized to other M-estimators lacking explicit forms of asymptotic normality, which has appeared in the literature, including, for example, \citet{Zeng2016}, \citet{Li2019} and \citet{Li2020}.

(iv) Finally, we approximate the two variance estimators with the least favorable direction estimated by the mean of covariates weighted by the estimated subject-specific rate. This further improves the computational efficiency of the variance estimators, as we shall demonstrate through simulation. We theoretically justify the approximated variance estimators when the time-varying covariates are constant within each observation interval, which is the case in YWP.

The remainder of the paper is organized as follows. After introducing the notation and assumptions, we present the estimation procedure under the Poisson working assumption and develop an EM algorithm in Section~\ref{sec:point_estimation}.  In Section~\ref{sec:asymptotic-properties}, we establish the asymptotic properties of the estimators without relying on the Poisson assumption, which confirms the robustness of the estimator. In Section~\ref{sec:varianceestimation}, we present the two variance estimators and show their consistency. In Section~\ref{sec:sim}, we demonstrate through simulation the robustness and superiority of the proposed method and variance estimation in practical situations. In Section~\ref{sec:app}, we analyze the YWP data using the proposed methods. We conclude the paper in Section~\ref{sec:discussion}  with a few brief remarks. Section~\ref{sec:proofasym} contains proofs of the asymptotic results along with variance estimation derivation details.

\section{Estimation Methodology \label{sec:point_estimation}}

Consider an event history study that consists of $n$ independent subjects.  For subject $i$, let $N_{i} (t)$ denote 
the total number of the occurrences of the recurrent event of interest up to time $t$ and suppose that 
$N_{i} (t)$ is observed only at the sequence of time points $T_{i0} = 0 < T_{i1} < \ldots < T_{iJ_{i}}$.  For subject $i$,
define the censoring indicator $\Delta_{i}\left(t\right)=I\left(T_{iJ_{i}}>t\right)$ and let $\bX_{i}\left(t\right)$ 
denote a $p$-dimensional vector of the possibly time-varying covariates associated with the subject.   The observed data have the form $\mathcal{O} = \{ O_{i} = (\{ N_{i}(T_{ij})\} _{j=1}^{J_{i}} ,
\bX_{i}(\cdot) , \{ T_{ij}\} _{j=1}^{J_{i}}) , i = 1 , \ldots , n \}$.  That is, only panel count data are
available.  

Define $\Delta N_{ij} = N_{i}\left(T_{ij}\right) - N_{i}\left(T_{i\left(j-1\right)}\right)$, the number of the occurrences of 
the recurrent event of interest between $T_{ij}$ and $T_{i\left(j-1\right)}$ in subject $i$, and assume that $N_{i}\left(t\right)$ 
satisfies model (\ref{eq:ratemodel}) and follows the non-homogeneous Poisson process, $j = 1 , ..., J_{i}$, $i = 1 , ..., n$. 
Then the observed likelihood function has the form
\begin{multline}
	L\left(\bb,\bL|\mathcal{O}\right)
	=\prod_{i=1}^{n}\prod_{j=1}^{J_{i}}\frac{1}{\Delta N_{ij}!}\exp\left(-\int_{T_{i\left(j-1\right)}}^{T_{ij}}\exp\left(\bb^{\top}\bX_{i}\left(t\right)\right)d\bL\left(t\right)\right)\\
	\times \left\{ \int_{T_{i\left(j-1\right)}}^{T_{ij}}\exp\left(\bb^{\top}\bX_{i}\left(t\right)\right)d\bL\left(t\right)\right\} ^{\Delta N_{ij}} \, , \label{eq:ll}
\end{multline}
and it is natural to maximize it for the estimation of $\bb$ and $\bL$.
However, since $\bX_{i}\left(t\right)$ is time-varying, and we do not know the exact form of $\bL\left(t\right)$,
the integral $\int_{T_{i\left(j-1\right)}}^{T_{ij}}\exp\left(\bb^{\top}\bX_{i}\left(t\right)\right)d\bL\left(t\right)$
cannot be easily calculated.  The use of numerical integration methods such as Gaussian quadrature would
require a certain degree of smoothness of the integrands and also the roughness of $\bX_{i}\left(t\right)$
could lead to numerical inaccuracy in integration.  In other words, the maximization of (\ref{eq:ll}) is complicated 
or not straightforward.  To overcome the challenge, we develop an EM algorithm, as inspired by \citet{Zeng2020}.

For the development of the EM algorithm and to make the integration in (\ref{eq:ll}) tractable, we adopt the
nonparametric approach by treating $\bL$ to be a step function with a nonnegative jump size
$\lambda_{k}$ at $t_{k}$ with $\lambda_{0} = 0$ and $0=t_{0}<t_{1}<\ldots<t_{K}$ denoting the ordered unique 
values of the $T_{ij}$'s for all $i$ and $j$. Let $\{ \, W_{ik} ; i = 1 , ... , n , k = 1 , ... , K \, \}$ be mutually independent 
Poisson random variables with means $\exp\left(\bb^{\top}\bX_{i}\left(t_{k}\right)\right)\lambda_{k}$. 
Then the likelihood function given in \eqref{eq:ll} can be rewritten as 
\begin{multline*}
	L\left(\bb,\bL|\mathcal{O}\right) = \prod_{i=1}^{n}\prod_{j=1}^{J_{i}}\frac{1}{\Delta N_{ij}!}\exp\left(-\sum_{T_{i\left(j-1\right)}<t_{k}\le T_{ij}}\exp\left(\bb^{\top}\bX_{i}\left(t_{k}\right)\right)\lambda_{k}\right) \nonumber \\
	 \times \left\{ \sum_{T_{i\left(j-1\right)}<t_{k}\le T_{ij}}\exp\left(\bb^{\top}\bX_{i}\left(t_{k}\right)\right)\lambda_{k}\right\} ^{\Delta N_{ij}} =\prod_{i=1}^{n}\prod_{j=1}^{J_{i}}\Pr\left(\sum_{_{T_{i\left(j-1\right)}<t_{k}\le T_{ij}}}W_{ik}=\Delta N_{ij}\right) \, . %
\end{multline*}
This suggests that one can augment the observed data by adding $W_{ik}$'s as the pseudo complete data
denoted by $\mathcal{C}$ for the development of the EM algorithm. 
It then follows that the pseudo complete-data log likelihood function has the form
\begin{equation}
	\ell\left(\bb,\bl|\mathcal{C}\right)=\sum_{i=1}^{n}\sum_{k=1}^{K}\Delta_{ik}\left\{ W_{ik}\left(\bb^{\top}\bX_{i}\left(t_{k}\right)+\log\left(\lambda_{k}\right)\right)-\exp\left(\bb^{\top}\bX_{i}\left(t_{k}\right)\right)\lambda_{k}\right\} \, ,\label{eq:ll_full}
\end{equation}
where $\bl=\left(\lambda_{1},\lambda_{2},\ldots,\text{\ensuremath{\lambda_{K}}}\right)$ and $\Delta_{ik}=\Delta_{i}\left(t_{k}\right)$.

In the E-step of the EM algorithm, we need to find the posterior expectation of $W_{ik}$
given the observed data and the estimators of $\bb$ and $\bl$ from the last iteration. 
Let $j_{ik}$ be the observation time index for subject $i$ such that $T_{i\left(j_{ik}-1\right)}<t_{k}\le T_{ij_{ik}}$
for any $k$ and $A_{ik}=\{ l:T_{i\left(j_{ik}-1\right)}<t_{l}\le T_{ij_{ik}}\} $.
We write the $q$th iteration of $\bb$ and $\bl$
as $\bb^{\left(q\right)}$ and $\bl^{(q)}=(\lambda_{1}^{(q)},\ldots,\lambda_{K}^{(q)})$.
In the $q$th iteration, the posterior mean of $W_{ik}$ given the observed data can be obtained as
\begin{align}
	\hat{E}\left[W_{ik}|\bb^{\left(q\right)},\bl^{\left(q\right)}\right] & =E\left[W_{ik}|\sum_{_{T_{i\left(j_{ik}-1\right)}<t_{k}\le T_{ij_{ik}}}}W_{ik}=\Delta N_{ij_{ik}},\bb^{\left(q\right)},\bl^{\left(q\right)}\right]\nonumber \\ 
	& =\frac{\Delta_{ik}\exp\left(\bb^{\left(q\right)\top}\bX_{i}\left(t_{k}\right)\right)\lambda_{k}^{\left(q\right)}}{\sum_{l\in A_{ik}}\Delta_{il}\exp\left(\bb^{\left(q\right)\top}\bX_{i}\left(t_{l}\right)\right)\lambda_{l}^{\left(q\right)}}\Delta N_{ij_{ik}}, \label{eq:EWdef}
\end{align}
because the conditional distribution of $W_{ik}$ is a binomial distribution
with the number of trials $\Delta N_{ij_{ik}}$ and the success probability
$E\left[W_{ik}\right]/\sum_{l\in A_{ik}}E\left[W_{lk}\right]$. 

In the M-step of the EM algorithm, it is straightforward to show that we can update $\bl$
by 
\begin{equation}
	\lambda_{k}^{\left(q+1\right)}=\sum_{i=1}^{n}\Delta_{ik}\hat{E}\left[W_{ik}|\bb^{\left(q\right)},\bl^{\left(q\right)}\right]/\left\{ \sum_{i=1}^{n}\Delta_{ik}\exp\left(\bb^{\left(q\right)\top}\bX_{i}\left(t_{k}\right)\right)\right\}  \, , \label{eq:lambdaupdate}
\end{equation}
$k=1,\ldots,K$. Plugging in (\ref{eq:lambdaupdate}) into the posterior expectation of (\ref{eq:ll_full}) given the observed data,
one can update $\bb$ by solving the following equation 
\[
\mathbf{U}\left(\bb|\bb^{\left(q\right)},\bl^{\left(q+1\right)}\right)=\sum_{i=1}^{n}\sum_{k=1}^{K}\Delta_{ik}\hat{E}\left[W_{ik}|\bb^{\left(q\right)},\bl^{\left(q+1\right)}\right]\left(\bX_{i}\left(t_{k}\right)-\bar{\bX}\left(t_{k};\bb\right)\right)=0.
\]
Here, 
$\bar{\bX}(t;\bb)=S^{\left(1\right)}\left(t,\bb\right)/S^{\left(0\right)}\left(t,\bb\right)$
with
$
S^{\left(u\right)}\left(t,\bb\right)=\frac{1}{n}\sum_{i=1}^{n}\bX_{i}\left(t\right)^{\otimes u}\Delta_{i}\left(t\right)\exp\left(\bb^{\top}\bX_{i}\left(t\right)\right)
$,
where $a^{\otimes0}=1$, $a^{\otimes1}=a$ and $a^{\otimes2}=aa^{\top}$
for a vector $a$ and $u=0,1,2$.  To reduce the computational burden,
one can update $\bb$ by using only the one-step Newton-Raphson at each M-step as 
$
\bb^{\left(q+1\right)}=\bb^{\left(q\right)}-\left\{\dot{\mathbf{U}}\left(\bb^{\left(q\right)}|\bb^{\left(q\right)},\bl^{\left(q+1\right)}\right)\right\}^{-1}\mathbf{U}\left(\bb^{\left(q\right)}|\bb^{\left(q\right)},\bl^{\left(q+1\right)}\right),
$
where 
\[
\dot{\mathbf{U}}\left(\bb|\bb^{\left(q\right)},\bl^{\left(q+1\right)}\right)=-\sum_{i=1}^{n}\sum_{k=1}^{K}\Delta_{ik}\hat{E}\left[W_{ik}|\bb^{\left(q\right)},\bl^{\left(q+1\right)}\right]\left\{ \frac{S^{\left(2\right)}\left(t_{k},\bb\right)}{S^{\left(0\right)}\left(t_{k},\bb\right)}-\bar{\bX}\left(t_{k};\bb\right)^{\otimes2}\right\} .
\]

For the implementation of the EM algorithm, it is apparent that one needs to choose a stopping or convergence
criterion, and a natural one, which is used in the numerical studies below, is when the sum of the relative differences of
two consecutive estimators of $\bb$ is smaller than a small positive constant such as $10^{-4}$.
It is worth pointing out that the EM algorithm developed above has several important computational advantages.
One is that it avoids the intractable integration in (\ref{eq:ll}),
which poses challenges under model (\ref{eq:ratemodel}) as discussed above.  Another is that
we only need to know the values of $\bX_{i}\left(t\right)$ at $t_{1},\ldots,t_{K}$
instead of the whole trajectories of covariates $\bX_{i}\left(t\right)$
over time, which are rarely available in practice. More importantly, the quantities from the last iteration of EM algorithms can serve as a cornerstone for variance estimation in Section \ref{sec:varianceestimation}.

\section{Asymptotic Theory\label{sec:asymptotic-properties}}

In this section, we establish the consistency, rate of convergence, and asymptotic normality of the estimators.
Let $\bt=(\bb,\bL)$ and $\hat{\bt} = (\hat{\bb} , \hat{\bL})$ denote the estimators
defined in the previous section with $\hat{\bL}\left(t\right)=\sum_{t_{k}<t}\hat{\lambda}_{k}$.
Assume that the observed data $\mathcal{O}=\left\{{O_i}\right\}_{i=1}^{n}$ are i.i.d. realizations of 
$O=\left(\left\{ \Delta N_{\cdot j}\right\} _{j=1}^{J},\bX,\left\{ T_{\cdot j}\right\} _{j=1}^{J}\right)$, and 
let $\mathfrak{B}_{p}$ and $\mathfrak{B}$ denote the collection of Borel sets in $\mathbb{R}^{p}$
and $\mathbb{R}$, respectively.   Let $\tau$ denote the maximum follow-up time for all
subjects, $\mathfrak{B}_{1}\left[0,\tau\right]=\left\{ B\cap\left[0,\tau\right]:B\in\mathfrak{B}\right\} $,
$\mathfrak{B}_{2}\left[0,\tau\right]=\mathfrak{B}_{1}\left[0,\tau\right]\times\mathfrak{B}_{1}\left[0,\tau\right]$,
and $\mathfrak{C}_{p}\left[0,\tau\right]$ denote the collection
of Borel sets in $\mathbb{C}_{p}=\left\{ f:\left[0,\tau\right]\to\mathbb{R}^{p}\right\} $.
Furthermore, for $B_{1}$, $B_{2}$, $B_{4}\in\mathfrak{B}_{1}\left[0,\tau\right]$ and 
$B_{3}\in\mathfrak{C}_{p}\left[0,\tau\right]$, define 
$\nu_{1}(B_{1}\times B_{2}\times B_{3})
=\int_{B_{3}}\sum_{k=1}^{\infty}\allowbreak\Pr(J=k|\bX=\bx)\sum_{j=1}^{k}\Pr(T_{\cdot(j-1)}\in B_{1},T_{\cdot j}\in B_{2}|J=k,\bX=\bx)d\mu_{\bX}(\bx)$,
$
\nu_{2}\left(B_{4}\times B_{3}\right)\allowbreak=\int_{B_{3}}\sum_{k=1}^{\infty}\allowbreak\Pr\left(J=k|\bX=\bx\right)\sum_{j=1}^{k}\Pr\left(T_{\cdot j}\in B_{4}|J=k,\bX=\bx\right)d\mu_{\bX}\left(\bx\right)
$
,
$
\nu_{3}\left(B_{1}\right)= \int_{\mathcal{X}}\sum_{k=1}^{\infty}\Pr\left(J=k|\bX=\bx\right)\allowbreak\sum_{j=1}^{k}\Pr\left(T_{\cdot J}\in B_{1}|J=k,\bX=\bx\right)d\mu_{\bX}\left(\bx\right)
$,
where $\mu_{\bX}\left(\bx\right)$ denotes the probability measure of $\bX$, $T_{\cdot j}$ denotes the population version of $T_{ij}$ and $\mathcal{X}$ is the sample space of $\bX$.
Also define $\mu_{1}\left(B_{1}\times B_{2}\right)=\nu_{1}\left(B_{1}\times B_{2}\times\mathcal{X}\right)$ and $\mu_{2}\left(B_{4}\right)=\nu_{2}\left(B_{4}\times \mathcal{X}\right)$.  

For the consistency and the rate of convergence, we will consider them under the $L_{2}$ metric 
$
d\left(\bt_{1},\bt_{2}\right)=\sqrt{\left\Vert \bb_{1}-\bb_{2}\right\Vert _{2}^{2}+\left\Vert \Delta\bL_{1} -
	\Delta\bL_{2}\right\Vert _{L_{2}\left(\mu_{1}\right)}^{2}} \, ,
$
where $\Delta\bL(T_{\cdot(j-1)},
T_{\cdot j})=\bL(T_{\cdot j})-\bL(T_{\cdot(j-1)})$.
Define $\mathcal{F}=\{ \bL:\bL$ is a nondecreasing function and $\bL\left(0\right)=0\} $,
and let $\lesssim$ ($\gtrsim$) represent the left (right) side bounded
by the right (left) side multiplied by a constant.  For the asymptotic properties of $\hat{\bt}$,
we need the following regularity conditions.

\begin{enumerate}[label=(C\arabic*) ]
	\item \label{enu:paraspace}The true value of $\bb$, denoted by $\bb_{0}$,
	lies in the interior of the compact set $\mathcal{B}\subset\mathbb{R}^{p}$.
	The true value of $\bL\in\mathcal{F}$, denoted by $\bL_{0}$,
	is continuously differentiable with positive and bounded first-order
	derivative $\bL_{0}^{\prime}$ in some interval $\left[\tau_{0},\tau\right]$
	where $\Pr\left(\cap_{j=1}^{J}\left\{ T_{\cdot j}\in\left[\tau_{0},\tau\right]\right\} \right)=1$
	and $\bL_{0}\left(\tau_{0}\right)>0$. 
	\item \label{enu:xbounded} With probability one, the covariate $\bX\left(t\right)$ of bounded variation over $[\tau_0,\tau]$,
	and its left limit exists for any $t$. %
	\item \label{enu:measure}The measure $\mu_{l}\times\mu_{\bX}$ is absolutely
	continuous with respect to $\nu_{l}$ for $l=1,2$. Furthermore, $\nu_{3}\left(\left\{ \tau\right\} \right)>0$.
	\item \label{enu:identifiability}If there exists a vector $\bb$ such
	that $\bb^{\top}\bX\left(t\right)=a\left(t\right)$ for some deterministic
	function $a\left(t\right)$ on $t\in\left[\tau_{0},\tau\right]$ almost surely (a.s.)
	with respect to $\nu_2$, then $\bb=0$ and $a\left(t\right)=0$ a.s. with respect to $\mu_2$.
	\item \label{enu:separable}The observation times are $s_{0}$-separated.
	That is, there exists a constant $s_{0}>0$, such that $\Pr(T_{\cdot j}-T_{\cdot(j-1)}\ge s_{0}|J,\bX)=1$
	for all $j=1,\ldots J$. 
	\item \label{enu:finite_obs}The number of observation times, $J$, is positive
	and finite with probability one.
	\item \label{enu:m0upperbound}The function $M_{0}\left(O\right)=\sum_{j=1}^{J}\Delta N_{\cdot j}\log\left\{ \Delta N_{\cdot j}\right\} $
	satisfies $E\left[M_{0}\left(O\right)\right]<0$. 
	\item \label{enu:expbound}For some $c_{0}\in\left(0,\infty\right)$, the
	function $\bX\to E\left[\exp\left(c_{0}N\left(\tau\right)\right)|\bX\left(s\right),0\le s\le\tau\right]$
	is uniformly bounded for $\bX\in\mathcal{X}$. 
	\item \label{enu:mu2con} The measure $\mu_{2}$ is absolutely continuous
	with respect to Lebesgue measure with a derivative $\dot{\mu}_{2}$
	that $\dot{\mu}_{2}\ge c_{1}>0$ for some positive constant $c_{1}$.
	\item \label{enu:L2inequal}For some $\rho\in\left(0,1\right)$, $a^{\top}\mathrm{Cov}\left(\bX\left(t\right),\bX\left(s\right)|U_{1},U_{2}\right)a\ge\rho a^{\top}E\left[\bX\left(t\right)\bX^{\top}\left(s\right)|U_{1},U_{2}\right]a$
	a.s. on $\nu_{1}$ for all $a\in\mathbb{R}^{p}$ and $t,s\in\left[\tau_{0},\tau\right]$
	where $\left(U_{1},U_{2},\bX\right)$ follows a distribution with density $\nu_{1}/\nu_{1}\left(\mathbb{R}^{+}\times\mathbb{R}^{+}\times\mathcal{X}\right)$. 
\end{enumerate}

Conditions \ref{enu:paraspace}-\ref{enu:xbounded} are
common assumptions for the true parameter $\bt_{0}=(\bb_{0},\bL_{0})$ and semi-parametric models with time-varying covariates, and in particular,
Condition \ref{enu:xbounded} is needed for boundedness and square integrability of the likelihood function. 
The first part of Condition \ref{enu:measure} together with \ref{enu:identifiability} and $\Pr(\cap_{j=1}^{J}\{ T_{\cdot j}\in[\tau_{0},\tau]\} )=1$ in \ref{enu:paraspace} ensures the identifiability of the parameters,
while the second part is a technical condition mainly used for showing
the boundedness of $\hat{\bL}\left(t\right)$ on $\left[0,\tau\right]$.
Condition \ref{enu:separable} means that the two consecutive
observation times should have at least $s_{0}$ difference, and Condition \ref{enu:finite_obs}
requires that the observation number is bounded.  Condition \ref{enu:m0upperbound}
holds when Condition \ref{enu:finite_obs} is true and the second moment of
$\Delta N_{\cdot j}$ is finite. Conditions \ref{enu:separable}-\ref{enu:m0upperbound}
are reasonable in practice, especially when the observation times
are discrete time points and the variance of $\Delta N_{\cdot j}$
is bounded. Conditions \ref{enu:expbound} and \ref{enu:L2inequal}
generalize the conditions C10 and C14 given in \citet{Wellner2007} to accommodate time-varying
covariates and could be justified similarly. Condition \ref{enu:mu2con}
is the same as the second part of the condition C12 given in \citet{Wellner2007}, implying
that the density of the observation times is strictly positive.
As shown below, 
Conditions \ref{enu:paraspace}-\ref{enu:m0upperbound} are for the consistency of the proposed estimators, 
and Conditions \ref{enu:expbound}-\ref{enu:L2inequal} are additional assumptions needed for 
establishing the convergence rate and asymptotic normality.  In the following, we
first establish the strong consistency and convergence rate of $\hat{\bt}$ and then 
the asymptotic normality of $\hat{\bb}$ with their proofs relegated to Section \ref{sec:proofasym}.

\begin{theorem}[Consistency]\label{thm:consistency}
	Suppose that Conditions \ref{enu:paraspace}-\ref{enu:m0upperbound}  and model (\ref{eq:ratemodel}) hold. 
Then we have that $d(\hat{\bt},\bt_{0})\to0$ a.s. as $n\to\infty$. 
\end{theorem}

\begin{theorem}[Rate of convergence]\label{thm:rate}
Suppose that Conditions \ref{enu:paraspace}-\ref{enu:L2inequal} and model (\ref{eq:ratemodel}) hold.
Then we have that $n^{1/3}d(\hat{\bt},\bt_{0})=O_{p}(1)$.
\end{theorem}

\begin{theorem}[Asymptotic normality]\label{thm:normality}
	Under the same conditions of Theorem~\ref{thm:rate}, we have that
$
\sqrt{n}(\hat{\bb}-\bb_{0})\to_{d}N\left(0,\bO^{-1}\bS(\bO^{-1})^{\top}\right),
$
where 
\[
\bO=-E\left[\sum_{j=1}^{J}\frac{\left\{ \int_{T_{\cdot\left(j-1\right)}}^{T_{\cdot j}}\exp\left(\bb_{0}^{\top}\bX\left(t\right)\right)\left(\bX\left(t\right)-\bar{\bx}\left(t\right)\right)d\bL_{0}\left(t\right)\right\} ^{\otimes2}}{\int_{T_{\cdot\left(j-1\right)}}^{T_{\cdot j}}\exp\left(\bb_{0}^{\top}\bX\left(t\right)\right)d\bL_{0}\left(t\right)}\right],
\]
and 
\begin{multline*}
	\bS=E\left[\left\{ \sum_{j=1}^{J}\left(\frac{\Delta N_{\cdot j}}{\int_{T_{\cdot\left(j-1\right)}}^{T_{\cdot j}}\exp\left(\bb_{0}^{\top}\bX\left(t\right)\right)d\bL_{0}\left(t\right)}-1\right)\right.\right.\\
	\times\left.\left.\int_{T_{\cdot\left(j-1\right)}}^{T_{\cdot j}}\exp\left(\bb_{0}^{\top}\bX\left(t\right)\right)\left(\bX\left(t\right)-\bar{\bx}\left(t\right)\right)d\bL_{0}\left(t\right)\right\} ^{\otimes2}\right].
\end{multline*}
Here, $\bar{\bx}\left(t\right) \in \mathbb{R}^{p}$ is some 
function on $\left[0,\tau\right]$ implicitly defined in Section \ref{sec:proofasym}.
Moreover, if the Poisson assumption
holds, $\hat{\bb}$ is asymptotically efficient.
\end{theorem}

We remark that both the strong consistency and convergence rate given in Theorems \ref{thm:consistency}-\ref{thm:rate} and
the asymptotic normality given in Theorem~\ref{thm:normality} are true as long as model \eqref{eq:ratemodel} holds 
and do not require the Poisson assumption.  In other words, the proposed estimator is robust.
The convergence rate $O(n^{1/3})$ is the typical optimal rate for semiparametric models \citep{Wellner2000,Wellner2007,Zeng2016,Zhao2017}.
Though it is slower than $n^{1/2}$, the common order of the convergence
rate from parametric MLEs, the asymptotic normality of $\hat{\bb}$ still follows by the convergence rate $\sqrt{n}$. 
In Theorem \ref{thm:normality}, the form of $\bS$ involves $\Delta N_{\cdot j}$ because
the explicit calculation of the expectation of $\bS$ requires
the formulation of $\mathrm{cov}\left(N\left(t\right),N\left(s\right)|\bX\left(s\right)\right)$
for $t\le s\in[0,\tau]$, which is unspecific in model (\ref{eq:ratemodel}).
If the Poisson assumption holds, $\bS$ would reduce to $-\bO$,
and both $\bS$ and $-\bO$ would be exactly the efficient information
matrix of $\hat{\bb}$, that is, $\hat{\bb}$ would be asymptotically efficient \citep{Murphy2000,Wellner2007}. 

It is apparent that for inference about $\bb$, one needs to estimate the asymptotic covariance of $\hat{\bb}$. 
Based on Theorem \ref{thm:normality}, it would be natural and typical to estimate $\bO$ and $\bS$ 
by their empirical counterparts. However, since  $\bar{\bx}\left(t\right)$ does not have an explicit form, no such empirical estimators would be easily available.
The profile likelihood approach would provide another choice but it would be sensitive 
to model specification.
Furthermore, its implementation would be computationally intensive as the lower-triangular elements in the asymptotic variance matrix are estimated one at a
time.
To address this, in the next section, we will present two consistent estimators of $\bO$ and $\bS$ that are free
of the Poisson assumption.
The proposed estimators will reuse the quantities from the last iteration of the EM algorithm
and thus can significantly reduce the computational burden. 

\section{Two Covariance Estimators of $\hat{\bb}$\label{sec:varianceestimation}}

We first briefly introduce the profile likelihood method and discuss what we hope to achieve in the variance estimation of $\hat{\bb}$.
\citet{Murphy1997} and \citet{Murphy2000} proposed the profile likelihood method based on the numerical derivatives of the profile log-likelihood function, which has been widely utilized in, for example, \citet{Su2016}, \citet{Zeng2016} and \citet{Zeng2020} with some modification.
The adaptation in \citet{Zeng2020} estimates the covariance matrix of $\hat{\bb}$ by $\hat{\bV}^{-1}/n$, where
$\hat{\bV}=\frac{1}{n}\sum_{i=1}^{n}\left[\{\partial\mathrm{p}\ell_{i}(\bb|\mathcal{O})/\partial\bb|_{\bb=\hat{\bb}}\} ^{\otimes2}\right]$.
Here, $\mathrm{p}\ell_{i}\left(\bb|\mathcal{O}\right)$ is the profile
log-likelihood function of $\bb$ by the $i$th subject
defined as $\ell_{i}(\bb,\hat{\bL}_{\bb}|\mathcal{O})$
where $\hat{\bL}_{\bb}=\arg\max_{\bL\in\mathcal{F}^{*}}\log L_{n}\left(\bb,\bL\right)$
with $\mathcal{F}^{*}$ being the set of step functions with nonnegative
jumps at $t_{k}$.
Given $\bb$, $\hat{\bL}_{\bb}$ can be obtained by the proposed EM algorithm through iterating \eqref{eq:lambdaupdate} only.
We can take $\hat{\bV}$ as the empirical covariance matrix
of the gradient of $\mathrm{p}\ell_{i}\left(\bb|\mathcal{O}\right)$
with respective $\bb$ evaluated at $\hat{\bb}$. The $l$th component
of the gradient is approximated by the first-order numerical derivative
of $\mathrm{p}\ell_{i}\left(\bb|\mathcal{O}\right)$ as 
$
\{\mathrm{p}\ell_{i}(\hat{\bb}+h_{n}e_{l}|\mathcal{O})-\mathrm{p}\ell_{i}(\hat{\bb}|\mathcal{O})\}/h_{n},
$
where $h_{n}=O(n^{-1/2})$, and $e_{l}$ is a $p$-dimensional
vector with the $l$th element being one and all others being zero.
This profile likelihood method tends to be
computationally inefficient as it essentially calculates a gradient in an element-wise fashion for each subject.
With large $p$ and sample size, the computational burden could be heavy.
Another drawback is the arbitrary choice of $h_{n}$.
Moreover, by Corollary 3 in \citet{Murphy2000}, $\hat{\bV}$ consistently estimates the asymptotic variance of $\hat{\bb}$ only when the Poisson model holds as \eqref{eq:ll} is constructed from the Poisson
process. To address these limitations, we put forward two new variance estimators based on Theorem \ref{thm:normality}. 

The first is a robust sandwich-type
variance estimator $\hat{\bO}^{-1}\hat{\bS}(\hat{\bO}^{-1})^{\top}/n$.
Here,
\[
\hat{\bO}=-\frac{1}{n}\sum_{i=1}^{n}\sum_{j=1}^{J}\frac{\left\{ \sum_{T_{i\left(j-1\right)}<t_{k}\le T_{ij}}\exp\left(\hat{\bb}^{\top}\bX_{i}\left(t_{k}\right)\right)\hat{\lambda}_{k}\left(\bX_{i}\left(t_{k}\right)-\hat{\bar{\bx}}_{k}\right)\right\} ^{\otimes2}}{\sum_{T_{i\left(j-1\right)}<t_{k}\le T_{ij}}\exp\left(\hat{\bb}^{\top}\bX_{i}\left(t_{k}\right)\right)\hat{\lambda}_{k}},
\]
and 
\[
\hat{\bS}=\frac{1}{n}\sum_{i=1}^{n}\left\{ \sum_{k=1}^{K}\Delta_{ik}\left(\hat{E}\left[W_{ik}|\hat{\bb},\hat{\bl}\right]-\exp\left(\hat{\bb}^{\top}\bX_{i}\left(t_{k}\right)\right)\hat{\lambda}_{k}\right)\left(\bX_{i}\left(t_{k}\right)-\hat{\bar{\bx}}_{k}\right)\right\} ^{\otimes2}.
\]
Here,
$\hat{\bar{\bx}}_{k}=-\left.d\log\left(\hat{\lambda}_{k}(\bb)\right)/d\bb\right|_{\bb=\hat{\bb}}$,
where $\hat{\bl}(\bb)=(\hat{\lambda}_{1}(\bb),\ldots,\hat{\lambda}_{K}(\bb))$ such that $\hat{\bL}_{\bb}\left(t\right)=\sum_{t_{k}<t}\hat{\lambda}_{k}\left(\bb\right)$.
As mentioned above, we can regard $\hat{\bO}$ and $\hat{\bS}$ as the respective
empirical counterparts and consistent estimators of $\bO$ and $\bS$, by rewriting
the integration with respect to $\hat{\bl}$ and by
replacing $\bar{\bx}\left(t_{k}\right)$ with its consistent estimator $\hat{\bar{\bx}}_{k}$.
This variance estimator is robust to the Poisson assumption as a direct
result of Theorem \ref{thm:normality}.
We can efficiently compute $l$th element of $\hat{\bar{\bx}}_{k}$ by $\log(\hat{\lambda}_{k}(\hat{\bb}+a_{n}e_{l}))-\log(\hat{\lambda}_{k}(\hat{\bb}))\} /a_{n}$ by iterating \eqref{eq:lambdaupdate} as we compute $\hat{\bV}$, where $a_n$ can be a small positive number depending on $n$ (e.g., $10^{-5}n^{-1/2}$). 
To prevent arithmetic overflow when $\hat{\lambda}_{k}(\hat{\bb})$ is very close or equal to zero, we set $\hat{\bar{\bx}}_{k}$ to zero in those cases.
This robust method has a much lighter computational burden than the profile likelihood method.
The terms $\hat{E}\left[W_{ik}|\hat{\bb},\hat{\bl}\right]$ and $\exp\left(\hat{\bb}^{\top}\bX_{i}\left(t_{k}\right)\right)\hat{\lambda}_{k}$ are immediately available to calculate $\hat{\bO}$ and $\hat{\bS}$
after finishing the EM algorithm.
By contrast, $\hat{\bV}$ requires computing the profile log-likelihood function of all subjects for $p$ different $\bb$.

For the second estimator, we first note that when the Poisson assumption holds, $\bO^{-1}\bS(\bO^{\top})^{-1}$ of the first variance estimator reduces
to $\bS^{-1}$, which coincides with the inverse of the efficient information
matrix of $\bb$. Then we can simply estimate the asymptotic variance
of $\hat{\bb}$ by $\hat{\bS}^{-1}/n$, which is 
further simplified computationally.
This efficient information-based estimator is closely connected to the profile likelihood method: when the Poisson
model holds, both $\hat{\bS}$ and $\hat{\bV}$ consistently estimate $\bS$, 
the variance of the efficient score function of $\hat{\bb}$ \citep{Murphy2000}.
In other words, $\hat{\bS}$ and $\hat{\bV}$ are asymptotically equal,
as confirmed by the simulation results in Section \ref{sec:sim}.
Therefore, the efficient information-based estimator $\hat{\bS}^{-1}/n$ provides a computationally efficient alternative to the profile likelihood estimator when the Poisson assumption could be justified.

We note that computational burden and complexity can be further reduced when \ref{enu:Xconst} holds.
\begin{enumerate}[label=(E\arabic*) ]
\item \label{enu:Xconst} With probability one, $\bX(t)$ is a piecewise-constant function of $t$ such that $\bX(t)$ is constant for $t$ within each internal $(T_{\cdot(j-1)},T_{\cdot j}]$, $j=1,\ldots,J$.
\end{enumerate}
Given \ref{enu:Xconst}, we can directly estimate $\bar{\bx}(t)$ directly from $\bar{\bX}(t;\hat{\bb})$, which is readily available after the EM algorithm as part of $\mathbf{U}(\hat{\bb}|\hat{\bt})$.
Thus, no additional computation and numerical differentiation are necessary.
In this case, $\hat{\bS}$ can be interpreted as the empirical variance of the contribution to $\mathbf{U}(\hat{\bb}|\hat{\bt})$ from each subject.
Condition \ref{enu:Xconst} is not uncommon in real-world studies, such as YWP. The time-varying covariates in YWP are considered constant for each subject within every observation interval (a quarter) because these covariates are collected as summary measures of the sexual behaviors of the subject since the last visit. Nonetheless, these covariates could still vary across different observation intervals. In addition, we note that $\bar{\bX}$ resembles $\bar{Z}$ in \citet{Lin2000} and thus the $\bar{\bX}$-based robust estimator is a generalized sandwich-type variance estimator, as proposed by \citet{Lin2000}, to panel count data. Importantly, even if Condition \ref{enu:Xconst} does not hold, the variance estimators based on $\bar{\bX}$ could still serve as a computational-efficient approximation to the original variance estimators.
In Section~\ref{sec:sim}, we shall demonstrate through simulation that the variance estimators using $\bar{\bX}$ closely resemble the original variance estimators even when \ref{enu:Xconst} is not satisfied.

Next, we prove the consistency of $\hat{\bar{\bx}}(t)=\sum_{k=1}^{K}I(t_{k-1}<t\le t_{k})\hat{\bar{\bx}}_{k}$.  Lemma~\ref{lem:xbarconsist} lays out the sufficient conditions for establishing the consistency of the proposed variance estimators. We first introduce additional notation for Lemma~\ref{lem:xbarconsist}. 
Let $\mathcal{H}_{3}=\{ h_{3}(t)\in\mathbf{R}^{p},t\in\left[0,\tau\right]:$ each element of $h_{3}(t)$ is of bounded variation over $\left[0,\tau\right]$$\}$. 
We prove the consistency of $\hat{\bar{\bx}}(t)$ under the metric for $\mathcal{H}_3$ $\tilde{d}(h_{31},h_{32})=\{ \int\Vert \int_{u_{2}}^{u_{1}}(h_{31}(t)-h_{32}(t))d\bL_{0}(t)\Vert _{2}^{2}d\nu_{1}(u_{1},u_{2},\bx)\} ^{1/2}$ for $h_{31},h_{32} \in \mathcal{H}_{3}$. 
\begin{lem}
	\label{lem:xbarconsist}
	If conditions \ref{enu:paraspace}-\ref{enu:L2inequal} and model	(\ref{eq:ratemodel}) hold,  $\tilde{d}\left(\hat{\bar{\bx}},\bar{\bx}\right) \to 0$ in probability as $n\to\infty$.
	Furthermore, if Condition \ref{enu:Xconst} additionally holds, $\bar{\bx}(t)$ reduce to $s^{\left(1\right)}(t,\bb_{0})/s^{\left(0\right)}(t,\bb_{0})$ and $\tilde{d}\left(\bar{\bX}(\cdot;\hat{\bb}),\bar{\bx}\right)\to0$, where $s^{\left(u\right)}\left(t,\bb\right)=E\left[\Delta(t)\bX\left(t\right)^{\otimes u}\exp(\bb^{\top}\bX\left(t\right))|T_{\cdot(j-1)},T_{\cdot j},J\right]$ for $u=0,1,2$.
\end{lem}
Lemma~\ref{lem:xbarconsist}  establishes the connection between the EM algorithm and the asymptotic normality in Theorem~\ref{thm:normality} since $\hat{\bar{\bx}}$ is based on $\hat{\bl}$  in EM algorithm.
Given Lemma~\ref{lem:xbarconsist}, we justify the aforementioned consistency of $\hat{\bO}$, $\hat{\bS}$ and $\hat{\bV}$ in the next theorem. 
\begin{theorem}\label{thm:varest}
	If conditions \ref{enu:paraspace}-\ref{enu:L2inequal} and model (\ref{eq:ratemodel}) hold, then we have that $\hat{\bO}\to_{p}\bO$,
and both $\hat{\bV}$ and $\hat{\bS}$ converge to $\bS$ in probability as $n\to\infty$. 
\end{theorem}

It is important to note that the conclusions of Lemma~\ref{lem:xbarconsist} and Theorem \ref{thm:varest} are not contingent on the Poisson assumption.
Thus, $\hat{\bV}$ and $\hat{\bS}$ are both consistent estimators for the variance of the efficient score function and hence asymptotically equivalent regardless of whether data come from a Poisson process. The profile likelihood estimator, on the other hand, may fail to yield accurate variance estimates for $\hat{\bb}$ when $\bS$ is not the asymptotic variance of $\hat{\bb}$ under a non-Poisson model. 
We elaborate the derivation of the variance estimation
methods and prove Theorem~\ref{thm:varest} in Section \ref{sec:proofvarest}.

\section{A Simulation Study \label{sec:sim}}

An extensive simulation study was conducted to evaluate the finite sample performance of the estimation approach 
proposed in the previous sections.  Herein, we considered the situation with two covariates $X_{1}(t)$ and $X_{2}$.
The time-varying covariate $X_{1}\left(t\right)$ was generated through
$
X_{1}\left(t\right)=\sum_{j=1}^{10}X_{1j}I(U_{(j-1)}<t\le U_{(j)})
$.
In the above, $\{X_{1j}\}_{j=1}^{10}$ were generated as independently and identically distributed (i.i.d.) random variables following the uniform distribution $\mathrm{Unif}\left(0,1\right)$. We then generated $\{U_{(j)}\}_{j=1}^{9}$ as an ordered i.i.d. sample from $\mathrm{Unif}\left(0,\tau\right)$, $U_{(0)}=0$ and $U_{(10)}=\tau$, where we set $\tau=2$ to represent the maximum follow-up time.
The covariate $X_{2}$ was assumed to be a constant and generated from $\mathrm{Unif}\left(0,1\right)$.

For the generation of panel count data, we considered two scenarios.  One is that the underlying recurrent event 
process $N_i (t)$ follows a non-homogeneous Poisson process with the rate function 
$\lambda (t) \, \exp( \beta_{1}X_{1}(t) + \beta_{2} X_{2})$.  The other is to assume that
$N_i (t)$ is a mixed non-homogeneous Poisson process with the rate function $ r \, \lambda\left(t\right) \, 
\exp\left(\beta_{1} X_{1}\left(t\right) + \beta_{2} X_{2}\right)$, where $r$ follows the Gamma distribution with both mean 
and  and variance being 1.  In the following, we set $\beta_{1} = 1$, $\beta_{2} = -1$ and 
$\lambda\left(t\right) = 16 / \left(1 + t\right)$ or $4 \, \left( \sin\left(4\pi t\right) + 2 \right)$.
For each subject, the total number of observations $J$ was generated from the zero-truncated Poisson distribution 
with mean $5$, and the observation times were taken to be the order statistics of a random sample of size $J$
from the uniform distribution over $\left(0,\bar{T}\right)$ with $\bar{T}\sim Unif\left(0.9\tau,\tau\right)$. 
The results given below are based on the sample size $n=100$, $200$ or $400$ with $1000$ replications.

Tables \ref{tab:simres_frac} and \ref{tab:simres_sin} present the results on estimation of the regression 
parameter $\bb$ given by the proposed estimation approach with $\lambda\left(t\right) = 16 / \left(1 + t\right)$ 
and $4 \, \left( \sin\left(4\pi t\right) + 2 \right)$, respectively.
In the tables, we calculated the average of the biases 
of the proposed estimates (Bias), the empirical standard error of the estimates (SE), the average of 
the proposed standard error estimates (SEE), and the 95\% empirical coverage probability (CP).
For the variance estimation, we considered two methods described in Section \ref{sec:varianceestimation}, 
the robust estimation (Robust) and the inverse of the estimated efficient information matrix (Info) with $a_{n}=10^{-5}n^{-1/2}$.
For comparison, we included the profile likelihood estimation (Profile), using $h_{n}=5n^{-1/2}$ as recommended by \citet{Zeng2016}.
In addition, we also computed Robust and Info with $\bar{\bx}_{k}$ being estimated by $\bar{\bX}(t_{k};\hat{\bb})$, whose SEE and CP are shown in parentheses in Tables \ref{tab:simres_frac} and \ref{tab:simres_sin}.
Since the setting of $X_{1}(t)$ does not satisfy Condition~\ref{enu:Xconst}, we assessed how well the $\bar{\bX}$-based variance estimators can approximate the original Robust and Info variance estimators.

The results suggest that the proposed estimator seems to be unbiased
and the robust variance estimation procedure appears to perform well, both independent of the Poisson assumption.
The normal approximation to the asymptotic distribution of the proposed estimator seems to be appropriate too,
and as expected, the results became better when the sample size increased.

It can be seen from the tables that under the Poisson assumption, all three variance estimation procedures yielded 
reasonable results and the coverage probabilities close to the 95\% nominal level.
However, when the Poisson assumption
was violated, both the Profile and Info methods severely underestimated the standard errors, leading to drastically reduced coverage probabilities.  In contrast, the Robust method was still able to provide reasonable estimates and maintain relatively 
satisfactory coverage probabilities, especially when the sample size was large.
In other words, the proposed robust variance estimator seems to be robust to the Poisson assumption. 
Another observation from the tables is that the Profile and Info methods gave almost identical results,
actually, further examination of the raw simulation results indicated that they were nearly the same for each replication.
This suggests that the Info method could be used as a computationally efficient substitute when the Poisson assumption holds.

One can further observe from the table that both SEE and CP of the original Robust and Info using numerical differentiation and those based on $\bar{\bX}$ are almost identical for all cases, even if $X_1{t}$ fluctuates on $[0,\tau]$ as a random piece-wise function of 10 pieces.
In Appendix~\ref{app:addsim}, we provide more simulation results with various settings where Condition~\ref{enu:Xconst} does not hold.
The results from these settings all show that the $\bar{\bX}$-based Robust and Info provide a good approximation to their original counterparts.

Then we assess the computational efficiency of Robust and Robust based on $\bar{\bX}$ compared with Profile.
We present the ratio of computational times of Robust and  $\bar{\bX}$-based Robust to those of Profile under different sample sizes in Appendix~\ref{app:addsim}. 
The results illustrate that,  compared with Profile, Robust is around 2 to 4 times faster, and $\bar{\bX}$-based Robust could be up to 100 times faster. 
In other words, the proposed robust variance estimators not only enjoy better estimation accuracy but also are much more computationally efficient than Profile.
Furthermore, the computational efficiency gain ratios of Robust and $\bar{\bX}$-based ones decrease when the sample size increases.
However, the computational efficiency gain ratios of Robust decrease slowly, and even with a very large sample size, e.g., 1600,   the computational efficiency of Robust roughly doubles that of Profile.
The computational efficiency ratio of $\bar{\bX}$-based Robust increases when $p$, the number of parameters, becomes larger, but that of Robust does not seem to vary with $p$. 
This suggests that $\bar{\bX}$-based Robust could be even more efficient when the number of covariates is large.

\begin{figure}[htpb]
	\centering{}\includegraphics[width=0.8\linewidth]{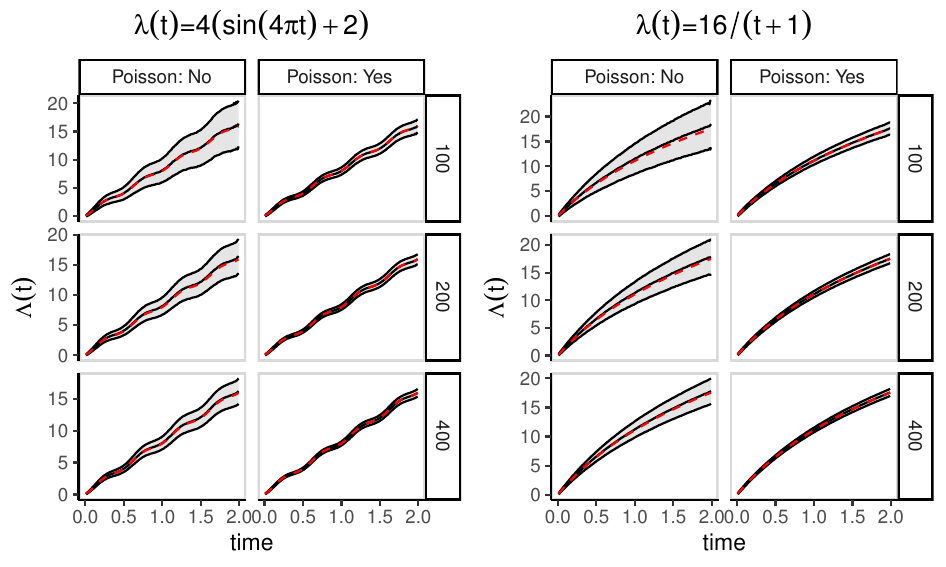}\caption{Estimates of the baseline cumulative rate function
		with $\lambda\left(t\right) = 16 / \left(1+t\right)$ and $\lambda\left(t\right) = 4 \left(\sin\left(4\pi t\right)+2\right)$: black solid lines --- the average of the estimates; red dashed lines ---
		the true $\bL\left(t\right)$; the upper and lower boundaries of the gray ribbons --- the mean plus and minus one pointwise
		empirical standard error of $\hat{\bL}\left(t\right)$, respectively.\label{fig:lamest}}
\end{figure}

\begin{table}[htpb]
	\centering{}\caption{Simulation results on the estimation of regression parameters with
		$\lambda\left(t\right)=16/\left(1+t\right)$. \label{tab:simres_frac}}
	
	\centering{}%
	\begin{tabular}{cccrrrrrrrr}
		\toprule 
		& Poisson  &  &  &  & \multicolumn{2}{c}{Robust ($\bar{\bX}$-based)} & \multicolumn{2}{c}{Info ($\bar{\bX}$-based)} & \multicolumn{2}{c}{Profile}\tabularnewline
		\cmidrule(lr){6-7}  \cmidrule(lr){8-9} \cmidrule(lr){10-11}
		$n$  & Model  &  & \mc{Bias}  & \mc{SE}  & \mc{SEE}  & \mc{CP}  & \mc{SEE}  & \mc{CP}  & \mc{SEE}  & \mc{CP}\tabularnewline
		\midrule 
		100  & Yes  & $\hat{\beta}_{1}$  & <0.001 & 0.092 & 0.088(0.088) & 92.9(92.8) & 0.094(0.094) & 94.9(94.9) & 0.094 & 94.5\tabularnewline
		&  & $\hat{\beta}_{2}$  & 0.001 & 0.082 & 0.082(0.082) & 94.9(94.9) & 0.087(0.087) & 96.1(96.1) & 0.087 & 95.9\tabularnewline
		& No  & $\hat{\beta}_{1}$  & -0.037 & 0.315 & 0.300(0.299) & 93.3(93.1) & 0.031(0.031) & 15.7(15.7) & 0.031 & 15.7\tabularnewline
		&  & $\hat{\beta}_{2}$  & -0.008 & 0.376 & 0.371(0.371) & 93.7(93.6) & 0.021(0.021) & 7.8(7.8) & 0.021 & 7.8\tabularnewline
		\addlinespace[3pt]
		200  & Yes  & $\hat{\beta}_{1}$  & -0.001 & 0.065 & 0.063(0.063) & 94.1(94.1) & 0.065(0.065) & 95.3(95.2) & 0.065 & 95.2\tabularnewline
		&  & $\hat{\beta}_{2}$  & <0.001 & 0.059 & 0.058(0.058) & 94.1(94.1) & 0.060(0.060) & 94.7(94.7) & 0.060 & 94.5\tabularnewline
		& No  & $\hat{\beta}_{1}$  & -0.017 & 0.221 & 0.211(0.211) & 92.5(92.5) & 0.021(0.021) & 14.8(14.8) & 0.021 & 14.9\tabularnewline
		&  & $\hat{\beta}_{2}$  & -0.003 & 0.271 & 0.263(0.263) & 92.5(92.5) & 0.014(0.014) & 8.4(8.4) & 0.014 & 8.4\tabularnewline
		\addlinespace[3pt]
		400  & Yes  & $\hat{\beta}_{1}$  & 0.002 & 0.045 & 0.045(0.045) & 94.0(94.0) & 0.045(0.045) & 94.4(94.4) & 0.045 & 94.5\tabularnewline
		&  & $\hat{\beta}_{2}$  & 0.001 & 0.041 & 0.041(0.041) & 94.7(94.7) & 0.042(0.042) & 94.8(94.8) & 0.042 & 94.8\tabularnewline
		& No  & $\hat{\beta}_{1}$  & -0.009 & 0.150 & 0.150(0.150) & 93.6(93.5) & 0.014(0.014) & 15.8(15.8) & 0.014 & 15.7\tabularnewline
		&  & $\hat{\beta}_{2}$  & -0.002 & 0.183 & 0.187(0.187) & 94.1(94.1) & 0.010(0.010) & 7.9(7.9) & 0.010 & 7.9\tabularnewline
		\bottomrule
	\end{tabular}
\end{table}
\begin{table}[htpb]
	\centering{}\caption{Simulation results on the estimation of regression parameters with
		$\lambda\left(t\right)=4\left(\sin\left(4\pi t\right)+2\right)$.
		\label{tab:simres_sin}}
	
	\centering{}%
	\begin{tabular}{cccrrrrrrrr}
		\toprule 
		& Poisson  &  &  &  & \multicolumn{2}{c}{Robust ($\bar{\bX}$-based)} & \multicolumn{2}{c}{Info ($\bar{\bX}$-based)} & \multicolumn{2}{c}{Profile}\tabularnewline
		\cmidrule(lr){6-7}  \cmidrule(lr){8-9} \cmidrule(lr){10-11}
		$n$  & Model  &  & \mc{Bias}  & \mc{SE}  & \mc{SEE}  & \mc{CP}  & \mc{SEE}  & \mc{CP}  & \mc{SEE}  & \mc{CP}\tabularnewline
		\midrule 
		100  & Yes  & $\hat{\beta}_{1}$  & -0.001 & 0.096 & 0.093(0.093) & 93.7(93.7) & 0.099(0.099) & 94.7(94.7) & 0.098 & 94.7\tabularnewline
		&  & $\hat{\beta}_{2}$  & 0.003 & 0.087 & 0.086(0.086) & 93.7(93.7) & 0.092(0.092) & 95.5(95.5) & 0.091 & 95.3\tabularnewline
		& No  & $\hat{\beta}_{1}$  & 0.003 & 0.299 & 0.291(0.290) & 93.3(93.3) & 0.034(0.034) & 18.7(18.7) & 0.034 & 18.5\tabularnewline
		&  & $\hat{\beta}_{2}$  & 0.010 & 0.384 & 0.368(0.367) & 92.2(92.2) & 0.023(0.023) & 8.6(8.6) & 0.023 & 8.6\tabularnewline
		\addlinespace[3pt]
		200  & Yes  & $\hat{\beta}_{1}$  & 0.001 & 0.068 & 0.066(0.066) & 94.4(94.4) & 0.068(0.068) & 94.7(94.7) & 0.068 & 94.8\tabularnewline
		&  & $\hat{\beta}_{2}$  & 0.003 & 0.062 & 0.061(0.061) & 94.0(94.0) & 0.063(0.063) & 95.4(95.4) & 0.063 & 95.5\tabularnewline
		& No  & $\hat{\beta}_{1}$  & -0.001 & 0.207 & 0.209(0.209) & 94.9(94.9) & 0.023(0.023) & 15.6(15.5) & 0.023 & 15.6\tabularnewline
		&  & $\hat{\beta}_{2}$  & 0.008 & 0.267 & 0.263(0.263) & 92.7(92.7) & 0.015(0.015) & 8.1(8.1) & 0.015 & 8.2\tabularnewline
		\addlinespace[3pt]
		400  & Yes  & $\hat{\beta}_{1}$  & <0.001 & 0.047 & 0.047(0.047) & 95.4(95.4) & 0.048(0.048) & 95.6(95.7) & 0.048 & 95.6\tabularnewline
		&  & $\hat{\beta}_{2}$  & 0.001 & 0.044 & 0.043(0.043) & 95.1(95.1) & 0.044(0.044) & 95.5(95.5) & 0.044 & 95.5\tabularnewline
		& No  & $\hat{\beta}_{1}$  & <0.001 & 0.145 & 0.147(0.147) & 95.0(95.0) & 0.016(0.016) & 14.4(14.4) & 0.016 & 14.4\tabularnewline
		&  & $\hat{\beta}_{2}$  & <0.001 & 0.189 & 0.186(0.186) & 94.5(94.5) & 0.010(0.010) & 7.3(7.3) & 0.011 & 7.3\tabularnewline
		\bottomrule
	\end{tabular}
\end{table}
To assess the performance of the proposed estimation procedure on the baseline cumulative rate function
$\bL (t)$, we plotted the averages of the proposed estimates $\hat{\bL}\left(t\right)$ corresponding to 
$\lambda\left(t\right)=16/\left(1+t\right)$ and $\lambda\left(t\right)=4\left(\sin\left(4\pi t\right)+2\right)$ in 
Figure \ref{fig:lamest}, respectively.  In addition, the true curves of $\bL (t)$ were
given too along with the confidence bands given by the average of the estimates plus and minus the pointwise 
empirical standard errors.   They suggest that the proposed estimator $\hat{\bL}\left(t\right)$ seems to be 
unbiased.  As expected, the estimator became more accurate when the sample size increased or
under the Poisson assumption.

\section{An Application \label{sec:app}}

We applied the proposed estimation procedure to the Young Women's Project 
(YWP) data mentioned in Section \ref{sec:intro}.  The study examined the infections with \emph{Chlamydia trachomatis} (CT), 
\emph{Neisseria gonorrhoeae} (NG), and \emph{Trichomonas vaginalis} (TV), which cause clinical diseases
of chlamydia, gonorrhea, and trichomoniasis, respectively. The study was conducted between 1999 and 2009,
and it consisted of young women aged 14 to 17 years.  Basic demographic and clinical information of the study participants was collected at enrollment. Sexual behaviors and infection were assessed at all quarterly visits and thus were time-varying. At these visits, participants underwent cervical and vaginal examinations; samples were collected for testing of STI with CT, NG, and TV.  The infected subjects were treated on the spot, so subsequent STI events represented new infections. The exact infection times were interval-censored, thus giving rise to panel count data on infection. Herein, we analyzed data from 271 participants that had completed at least one follow-up visit. We focused on the effects of the seven social, familial, and behavioral risk factors on each of the three types of STIs.
The seven risk factors were
\begin{enumerate}
	\item $X_{1}$: age at enrollment, 
	\item $X_{2}(t)$: 1 when having zero unprotected sexes in
	the previous quarter; 0 otherwise, 
	\item $X_{3}(t)$: 1 when having five or more times' unprotected
	sex in the previous quarter; 0 otherwise, 
	\item $X_{4}(t)$: 1 when having multiple sex partners in the
	previous quarter; 0 otherwise, 
	\item $X_{5}$: 1 when living with parents at the time of enrollment; 0 otherwise, 
	\item $X_{6}$: 1 when having used marijuana in the past three
	months prior to enrollment; 0 otherwise, and 
	\item $X_{7}$: 1 when having received sex education from parents
	prior to enrollment; 0 otherwise.
\end{enumerate}
Among the seven factors, $X_{2}$, $X_{3}$ and $X_{4}$ were considered time-varying, and the rest time-invariant.

Table \ref{tab:realdata} presents the analytical results produced by the proposed method, with the application of three variance estimation procedures. As stated earlier, all the time-varying factors satisfy Condition~\ref{enu:Xconst} because the summaries of sexual behaviors since the last visit were constant between two consecutive visits. In addition, the $\bar{\bX}$-based variance estimators were identical to the original variance estimators up to 6 digits, further strengthening the conclusion of Lemma~\ref{lem:xbarconsist}.
Hence, we only presented the results from the $\bar{\bX}$-based variance estimators in Table \ref{tab:realdata}. For comparison, we also applied the estimating equation-based method proposed in \citep{Hu2003} under the proportional mean model \eqref{eq:meanmodel}.
The results based on the robust variance estimation suggested that having more than one sexual partner was associated with significantly increased risk for all three types of STIs. The finding is consistent with the existing medical literature \citep{Batteiger2010}. The analysis also indicated that having no unprotected sex was significantly associated with the decreased risk of the infection with NG or TV but did not have a significant effect on the CT infection. On the other hand, having five or more times unprotected sex in the previous quarter had no significant association with the infection risk of all three organisms. Among the time-invariant factors, living with parents was associated with reduced CT and TV infection risks, while drug use and the received sex education were associated with increased CT risk but not NG and TV risks.

As seen in the simulation study, the profile likelihood and efficient information variance methods gave nearly identical results, which are generally similar to those given by the robust variance method, except that the two former methods also indicate a negative association between age and CT acquisition, again consistent with the current scientific understanding \citep{Torrone2014,Li2015}.  A possible explanation for this is that the immaturity of the host immune response makes younger women more susceptible to CT infection \citep{Batteiger2010}. Another observation is that having five or more times' unprotected sex in the previous quarter tended to attenuate the risk for TV. Such findings should be viewed with caution since the two methods could underestimate the variance, as seen in the simulation study.   In comparison, one can also see from Table \ref{tab:realdata} that the proportional mean model method failed to detect the influence of two factors, multiple partners and zero unprotected sex. We also obtained the estimated cumulative baseline rate functions corresponding to each and plotted them in Appendix~\ref{app:reallam}.

\begin{table}
	\caption{Analysis results of Young Women's Project.\label{tab:realdata}}
	\begin{centering}
		\begin{tabular}{@{}lrcrcrcrcrcr@{}}
			\toprule 
			& \multicolumn{7}{c}{Proposed Method} &  & \multicolumn{3}{c}{Method in }\tabularnewline
			\cmidrule{2-8}  
			&  & \multicolumn{2}{c}{Robust} & \multicolumn{2}{c}{Profile} & \multicolumn{2}{c}{Info} &  & \multicolumn{3}{c}{Hu et al. (2003)}\tabularnewline
			\cmidrule(lr){3-4} \cmidrule(lr){5-6} \cmidrule(lr){7-8}  \cmidrule(lr){10-12}  
			& \mc{Est} & \mc{SE} & \mc{$p$-value} & \mc{SE} & \mc{$p$-value} & \mc{SE} & \mc{$p$-value} &  & \mc{Est} & \mc{SE} & \multicolumn{1}{c@{}}{$p$-value}\tabularnewline
			\cmidrule{1-8}\cmidrule{10-12}
			\multicolumn{12}{c}{CT}\tabularnewline
			$X_{1}$ & -0.053  & 0.038  & 0.164  & 0.026  & 0.040  & 0.026  & 0.039  &  & -0.072  & 0.039  & 0.066\tabularnewline
			$X_{2}$ & -0.251  & 0.137  & 0.068  & 0.157  & 0.109  & 0.157  & 0.110  &  & -0.185  & 0.113  & 0.102\tabularnewline
			$X_{3}$ & -0.152  & 0.137  & 0.267  & 0.159  & 0.340  & 0.159  & 0.338  &  & -0.088  & 0.079  & 0.267\tabularnewline
			$X_{4}$ & 0.651  & 0.137  & <0.001  & 0.133  & <0.001  & 0.133  & <0.001  &  & 0.234  & 0.129  & 0.070\tabularnewline
			$X_{5}$ & -0.455  & 0.214  & 0.034  & 0.161  & 0.005  & 0.162  & 0.005  &  & -0.417  & 0.164  & 0.011\tabularnewline
			$X_{6}$ & 0.811  & 0.332  & 0.015  & 0.319  & 0.011  & 0.319  & 0.011  &  & 1.136  & 0.473  & 0.016\tabularnewline
			$X_{7}$ & -0.483  & 0.246  & 0.049  & 0.168  & 0.004  & 0.168  & 0.004  &  & -0.409  & 0.201  & 0.042\tabularnewline
			\addlinespace[3pt]
			\multicolumn{12}{c}{NG}\tabularnewline
			$X_{1}$ & 0.017  & 0.053  & 0.745  & 0.041  & 0.675  & 0.041  & 0.676  &  & 0.017  & 0.054  & 0.759\tabularnewline
			$X_{2}$ & -0.479  & 0.229  & 0.037  & 0.214  & 0.025  & 0.212  & 0.024  &  & -0.116  & 0.147  & 0.431\tabularnewline
			$X_{3}$ & -0.219  & 0.217  & 0.313  & 0.188  & 0.244  & 0.188  & 0.244  &  & 0.026  & 0.120  & 0.832\tabularnewline
			$X_{4}$ & 0.966  & 0.189  & <0.001  & 0.184  & <0.001  & 0.183  & <0.001  &  & 0.544  & 0.117  & <0.001\tabularnewline
			$X_{5}$ & -0.240  & 0.391  & 0.538  & 0.279  & 0.389  & 0.278  & 0.388  &  & -0.365  & 0.350  & 0.298\tabularnewline
			$X_{6}$ & 0.599  & 0.545  & 0.272  & 0.400  & 0.134  & 0.399  & 0.133  &  & 1.335  & 0.854  & 0.118\tabularnewline
			$X_{7}$ & -0.203  & 0.658  & 0.758  & 0.196  & 0.300  & 0.196  & 0.300  &  & -0.506  & 0.485  & 0.297\tabularnewline
		    \addlinespace[3pt]
			\multicolumn{12}{c}{TV}\tabularnewline
			$X_{1}$ & 0.034  & 0.049  & 0.488  & 0.031  & 0.270  & 0.031  & 0.267  &  & 0.024  & 0.057  & 0.677\tabularnewline
			$X_{2}$ & -0.469  & 0.201  & 0.020  & 0.167  & 0.005  & 0.168  & 0.005  &  & -0.354  & 0.146  & 0.015\tabularnewline
			$X_{3}$ & -0.335  & 0.187  & 0.073  & 0.157  & 0.033  & 0.158  & 0.034  &  & -0.017  & 0.111  & 0.876\tabularnewline
			$X_{4}$ & 0.860  & 0.180  & <0.001  & 0.139  & <0.001  & 0.139  & <0.001  &  & 0.396  & 0.154  & 0.010\tabularnewline
			$X_{5}$ & -0.684  & 0.256  & 0.007  & 0.227  & 0.003  & 0.231  & 0.003  &  & -0.661  & 0.284  & 0.020\tabularnewline
			$X_{6}$ & -0.079  & 0.347  & 0.820  & 0.251  & 0.753  & 0.252  & 0.754  &  & 0.343  & 0.459  & 0.455\tabularnewline
			$X_{7}$ & -0.181  & 0.364  & 0.619  & 0.257  & 0.482  & 0.258  & 0.484  &  & 0.021  & 0.280  & 0.940\tabularnewline
			\bottomrule
		\end{tabular}
		\par\end{centering}
\end{table}

\section{Discussion and Concluding Remarks \label{sec:discussion}}

This paper presents a regression model for panel count data, considering time-varying covariates, and proposes a robust likelihood-based estimation procedure within the proportional rate modeling framework. While the alternative proportional means model has been used in the literature, maintaining the monotonicity of the mean function is often difficult, if not impossible. Interestingly, in the analysis of YWP data, the mean model failed to detect the effects of known STI risk factors (multiple partners and practice of unprotected sex), thus raising questions about its applicability in practical situations. This research has attempted to address several key challenges that analysts face when using proportional rates models with time-varying covariates, namely the robustness of estimators when the Poisson process assumption is violated and the computational issues associated with the profile likelihood method for variance estimation. We present a robust solution with carefully developed asymptotic properties of the proposed estimators that do not rely on the Poisson process assumption. Through numerical studies, we have demonstrated the good performance of the proposed procedures in simulated situations.

One significant contribution of the proposed approach is the introduction of a variance estimation method for $\hat{\bb}$ based on the EM algorithm. This innovation has played an essential role in overcoming the challenge arising from the absence of an explicit form for the asymptotic variance. The solution presented in this paper not only addresses the specific difficulty in panel count data analysis but also sheds light on the broader issue of handling the lack of closed-form expressions for asymptotic variance. This challenge is commonly encountered in the analysis of interval-censored failure time data under semiparametric models, where EM algorithms are frequently employed for estimation. In such situations, analysts typically resort to computationally heavy methods for variance estimation, such as profile likelihood or bootstrapping methods  \citep{Gao2019, Li2019, Li2020}. An important justification for the proposed variance estimation technique is that consistent estimation of $\bar{\bx}$ is achievable through the derivatives of the discretized profiled rate function in the EM algorithm. Furthermore, a critical observation is that the efficient score function at $\bt_0$ is econnected to its EM counterpart. Consequently, the variance estimation methods utilized in this work have the potential to be applied or extended to other models where profile likelihood methods are employed. We believe that this research has laid the initial theoretical foundation for future extensions in this direction.

As highlighted earlier, this research holds promising potential for extension in various directions. For instance, future work could encompass the analysis of mixed recurrent event and panel count data \citep{Zhu2014, Yu2017, Ge2020}, modeling multivariate panel count data, and the development of robust goodness-of-fit tests. With these extensions in mind, we present this study as an initial yet significant stepping stone toward addressing a broader range of methodological issues concerning panel count data.

\section*{Acknowledgments}
We would like to express our sincere appreciation to two anonymous reviewers, an Associate Editor and the Editor for their invaluable feedback and constructive suggestions that significantly improved the quality of this paper.
This work is partially supported by NIH grants HL095086 and AA026969.

\appendix

\section{Proofs of the Asymptotic Properties and Variance Estimation}\label{sec:proofasym}

In this section, we will sketch the proofs for Theorems \ref{thm:consistency} - \ref{thm:normality} and Theorem \ref{thm:varest}
as well as providing some details about the derivation of the proposed variance estimation.
The proofs of the asymptotic properties rely on the modern empirical process theory.
Let $\mathbf{\mathbb{P}}_{n}$ and $\mathbf{P}$ denote the empirical
measure and the true probability measure, respectively.
Denote the empirical
process by $\mathbb{G}_{n}=\sqrt{n}\left(\mathbb{P}_{n}-\mathbf{P}\right)$.
For convenience, denote $D_{j}^{0}(\bt)=\int_{T_{\cdot(j-1)}}^{T_{\cdot j}}\exp(\bb^{\top}\bX(t))d\bL(t)$ and $D_{j}^{1}(\bt,h)=\int_{T_{\cdot(j-1)}}^{T_{\cdot j}}\exp(\bb^{\top}\bX(t))h(t)d\bL(t)$ where $h(t)$ is scalar- or vector-value function of $t \in [\tau_0,\tau]$.
Let
$\mathbb{M}_{n}\left(\bt\right)=n^{-1}\ell\left(\bb,\bL\right)=\mathbf{\mathbb{P}}_{n}m_{\bt}$
and $\mathbf{M}\left(\bt\right)=\mathbf{P}m_{\bt}$, where $m_{\bt}\left(O\right)=\sum_{j=1}^{J}\Delta N_{\cdot j}\log\{ D_{j}^{0}(\bt)\} -D_{j}^{0}(\bt)$.
Here, we omit the term $-\log\left(\Delta N_{ij}!\right)$
as it does not involve the parameters.
For simplicity, we suppress $(O)$ in $m_{\bt}\left(O\right)$ in the following.

\subsection{Proof of Theorem \ref{thm:consistency}}

We begin with proving that the true parameter $\bt_{0}=\left(\bb_{0},\bL_{0}\right)$
is the unique maximizer of $\mathbf{M}\left(\bt\right)$. From
the definition of $\mathbf{M}\left(\bt\right)$, 
\begin{equation}
	\mathbf{M}\left(\bt_{0}\right)-\mathbf{M}\left(\bt\right)=\int\int_{u_{1}}^{u_{2}}\exp\left(\bb^{\top}\bx\left(t\right)\right)d\bL\left(t\right)h\left[\frac{\int_{u_{1}}^{u_{2}}\exp\left(\bb_{0}^{\top}\bx\left(t\right)\right)d\bL_{0}\left(t\right)}{\int_{u_{1}}^{u_{2}}\exp\left(\bb^{\top}\bx\left(t\right)\right)d\bL\left(t\right)}\right]d\nu_{1}\left(u_{1},u_{2},\bx\right),
	\label{eq:Mtheta0_m_Mtheta}
\end{equation}
where $h\left(z\right)=z\log z-z+1$. The function $h\left(z\right)$
is nonnegative for all $z>0$ with the equality holding only at $z=1$.
Therefore, $\mathbf{M}\left(\bt_{0}\right)\ge\mathbf{M}\left(\bt\right)$
and $\mathbf{M}\left(\bt_{0}\right)=\mathbf{M}\left(\bt\right)$
if and only if 
\begin{equation}
	\int_{u_{1}}^{u_{2}}\exp\left(\bb_{0}^{\top}\bx\left(t\right)\right)d\bL_{0}\left(t\right)=\int_{u_{1}}^{u_{2}}\exp\left(\bb^{\top}\bx\left(t\right)\right)d\bL\left(t\right),\quad \text{a.e. with respect to }\nu_{1}. \label{eq:expbXdLam_equal}
\end{equation}
Under condition \ref{enu:paraspace}, (\ref{eq:expbXdLam_equal}) reduces to
$
\exp(\bb_{0}^{\top}\bx(t))\bL_{0}^{\prime}(t)=\exp(\bb^{\top}\bx(t))\bL^{\prime}(t),
$
a.e. with respect to $\nu_{2}$. By the similar argument to prove (5.2) in \citet[page 2123]{Wellner2007},
we further have 
$
\{ 1-\exp((\bb_{0}-\bb)^{\top}\bx(t))\} ^{2}=(\bL^{\prime}(t)/\bL_{0}^{\prime}(t)-1)^{2} 
$
a.e. with respect to $\nu_{2}$.
This leads to $\bb_{0}=\bb$
and $\bL_{0}^{\prime}\left(t\right)=\bL^{\prime}\left(t\right)$ a.e. with respect to $\mu_{2}$
in view of \ref{enu:measure} and \ref{enu:identifiability}.
The latter equality
also implies $\bL_{0}\left(t\right)=\bL\left(t\right)$ a.e.
with respect to $\mu_{2}$ since $\Lambda(0)=\Lambda_{0}(0)=0$.
It follows that $\bt_{0}$
is the unique maximizer of $\mathbf{M}\left(\bt\right)$.

We next show $\hat{\bL}\left(t\right)$ is uniformly
bounded for $t\in\left[0,\tau\right]$ a.s..
For any given
$\epsilon>0$, let $\tilde{\bt}_{\epsilon}=(\hat{\bb},(1-\epsilon)\hat{\bL}+\epsilon\bL_{0})=\hat{\bt}+\epsilon(0,\bL_{0}-\hat{\bL}_{n})$.
Since $\hat{\bt}$ maximizes $\mathbb{M}_{n}(\cdot)$,
we have $\mathbb{M}_{n}(\hat{\bt})\ge\mathbb{M}_{n}(\tilde{\bt}_{\epsilon})=\mathbb{M}_{n}(\hat{\bt}+\epsilon(0,\bL_{0}-\hat{\bL}_{n}))$.
Therefore, 
$
	0\ge  \lim_{\epsilon\to0}\{ \mathbb{M}_{n}(\hat{\bt}+\epsilon(0,\bL_{0}-\hat{\bL}_{n}))-\mathbb{M}_{n}(\hat{\bt})\} /\epsilon=G_{n}-\mathbb{P}_{n}(\sum_{j=1}^{J}D_{j}^{0}((\hat{\bb},\bL_{0}))+\Delta N_{\cdot j}),
$
where $G_{n}=\mathbb{P}_{n}\sum_{j=1}^{J}\Delta N_{\cdot j}D_{j}^{0}((\hat{\bb},\bL_{0}))/D_{j}^{0}(\hat{\bt})+D_{j}^{0}(\hat{\bt})$.
Then,
\begin{equation}
	G_{n}\le C_{1}\mathbb{P}_{n}\left\{ \sum_{j=1}^{J}\Delta N_{\cdot j}+\bL_{0}\left(T_{\cdot J}\right)\right\} \to_{\text{a.s.}}C_{1}\mathbf{P}\left\{ \sum_{j=1}^{J}\Delta N_{\cdot j}+\bL_{0}\left(T_{\cdot J}\right)\right\} ,\label{eq:PNJfnite}
\end{equation}
for some finite $C_{1}$ by conditions \ref{enu:paraspace},
\ref{enu:xbounded} and \ref{enu:finite_obs} and the strong law of
large numbers. The limit on the right hand of \eqref{eq:PNJfnite}
is finite under conditions \ref{enu:paraspace}, \ref{enu:xbounded}
and \ref{enu:finite_obs}. On the other hand, 
$
\limsup_{n\to\infty}G_{n}\ge\limsup_{n\to\infty}\mathbb{P}_{n}\sum_{j=1}^{J}D_{j}^{0}(\hat{\bt}),
$
because $\Delta N_{\cdot j}$ and $D_{j}^{0}(\bt)$
are both nonnegative by \ref{enu:paraspace} and \ref{enu:separable}.
Under conditions \ref{enu:paraspace}, \ref{enu:xbounded} and \ref{enu:finite_obs},
for $0<\tau_{0}<\tau$, 
$
\lim\sup_{n\to\infty}\mathbb{P}_{n}\sum_{j=1}^{J}D_{j}^{0}(\hat{\bt})  \gtrsim\lim\sup_{n\to\infty}\mathbb{P}_{n}I\left\{ \tau_{0}\le T_{\cdot J}\le\tau\right\} \hat{\bL}\left(T_{\cdot J}\right)\ge\lim\sup_{n\to\infty}\hat{\bL}\left(\tau\right)\nu_{3}\left(\left\{ \tau\right\} \right).
$
The right hand side of the last equality is obviously smaller than
the left hand of \eqref{eq:PNJfnite}.
Hence, $\hat{\bL}\left(\tau\right)\nu_{3}\left(\left\{ \tau\right\} \right)$
is finite with probability one.
Consequently, $\hat{\bL}\left(t\right)$ is uniformly bounded
a.s. $t\in\left[0,\tau\right]$ because  $\nu_{3}\left(\left\{ \tau\right\} \right)>0$
in \ref{enu:measure}.

Finally, we prove the strong consistency of $\hat{\bt}$ by the Helly
selection theorem and one-sided Glivenko-Cantelli theorem stated as Theorem A.1 in \citet{Wellner2000}.
Using the arguments similar to prove Theorem 4.1 of \citet{Wellner2000} and those on page 2125 of \citet{Wellner2007}, we obtain the strong consistency of $\hat{\bb}$ and the almost everywhere convergence
of $\hat{\bL}\left(t\right)$ to $\bL_{0}\left(t\right)$
on $\mu_{2}$ almost surely.
Due to the uniform boundedness
of $\hat{\bL}$, the dominated convergence theorem yields the
strong consistency of $\hat{\bt}$ in the metric $d\left(\cdot,\cdot\right)$.

\subsection{Proof of Theorem~\ref{thm:rate}}

We verify the conditions of Theorem 3.2.5 of \citet{Vaart1996} to derive the
convergence rate of $\hat{\bt}$.
We first show $\mathbf{M}\left(\bt_{0}\right)-\mathbf{M}\left(\bt\right)\gtrsim d^{2}\left(\bt,\bt_{0}\right)$.
From \eqref{eq:Mtheta0_m_Mtheta}, the relation between $d^{2}\left(\bt,\bt_{0}\right)$ and the lower bound of $\mathbf{M}\left(\bt_{0}\right)-\mathbf{M}\left(\bt\right)$ is mainly determined by $h\left(z\right)=z\log z-z+1$. Since $h\left(z\right)\ge\left(1/4\right)\left(z-1\right)^{2}$ for
$z$ in a small enough neighborhood of $1$, for any $\bt$ in
a sufficiently small neighborhood of $\bt_{0}$, by \ref{enu:paraspace} and \ref{enu:xbounded}, $\mathbf{M}(\bt_{0})-\mathbf{M}(\bt)\gtrsim  \int\{ \int_{u_{1}}^{u_{2}}\exp(\bb_{0}^{\top}\bx(t))d\bL_{0}(t)-\int_{u_{1}}^{u_{2}}\exp(\bb^{\top}\bx(t))d\bL(t)\} ^{2}d\nu_{1}(u_{1},u_{2},\bx)
$.
Let $g\left(\xi\right)=\int I\left(U_{1}<t<U_{2}\right)\exp(\bb_{\xi}^{\top}\bX\left(t\right))d\bL_{\xi}\left(t\right)$
where $\bb_{\xi}=\xi\bb+\left(1-\xi\right)\bb_{0}$ and $\bL_{\xi}=\xi\bL+\left(1-\xi\right)\bL_{0}$.
By the mean value theorem, there exists $\xi^{*}\in\left[0,1\right]$ such that $\mathbf{M}\left(\bt_{0}\right)-\mathbf{M}\left(\bt\right) \gtrsim \nu_{1}\left\{ g_{1}h_{1}+g_{2}\right\} ^{2}$
where $g_{1}\left(U_{1},U_{2},\bX\right)=\int_{U_{1}}^{U_{2}}\left(\bb-\bb_{0}\right)^{\top}\bX\left(t\right)d\bL_{0}\left(t\right)$,
$g_{2}\left(U_{1},U_{2}\right)=\bL\left(U_{2}\right)-\bL_{0}\left(U_{2}\right)-\left(\bL\left(U_{1}\right)-\bL_{0}\left(U_{1}\right)\right)$
and 
$
h_{1}(U_{1},U_{2},\bX)=\xi\int_{U_{1}}^{U_{2}}(\bb-\bb_{0})^{\top}\bX(t)d\bL(t)\allowbreak/\int_{U_{1}}^{U_{2}}(\bb-\bb_{0})^{\top}\bX(t)d\bL_{0}(t)+1-\xi.
$
For convenience, we write
the expectation under $\nu_{1}$ as $E_{1}$.
To apply Lemma 8.8 in \citet{Vaart2002}, we need to bound  
$
\left\{ E_{1}\left[g_{1}g_{2}\right]\right\} ^{2}
$ by $\left(1-\eta\right)E_{1}\left[g_{1}^{2}\right]E_{1}\left[g_{2}^{2}\right]$ for some $\eta\in(0,1)$.
By the Cauchy--Schwarz
inequality, 
$
	 \{ E_{1}[g_{1}g_{2}]\} ^{2} =  \{ E_{1}[g_{2}E_{1}[g_{1}|U_{1},U_{2}]]\} ^{2}	\le  E_{1}[g_{2}^{2}]E_{1}[\{ E_{1}[g_{1}|U_{1},U_{2}]\} ^{2}]
$.
Furthermore, 
\begin{align*}
	& \left\{ E_{1}\left[g_{1}|U_{1},U_{2}\right]\right\} ^{2}\\
	= & \int_{U_{1}}^{U_{2}}\int_{U_{1}}^{U_{2}}\left(\bb-\bb_{0}\right)^{\top}E_{1}\left[\bX\left(t\right)|U_{1},U_{2}\right]E_{1}\left[\bX^{\top}\left(s\right)|U_{1},U_{2}\right]\left(\bb-\bb_{0}\right)d\bL_{0}\left(t\right)d\bL_{0}\left(s\right)\\
	\le & \left(1-\rho\right)\int_{U_{1}}^{U_{2}}\int_{U_{1}}^{U_{2}}\left(\bb-\bb_{0}\right)^{\top}E_{1}\left[\bX\left(t\right)\bX^{\top}\left(s\right)|U_{1},U_{2}\right]\left(\bb-\bb_{0}\right)d\bL_{0}\left(t\right)d\bL_{0}\left(s\right)\\
	= & \left(1-\rho\right)E_{1}\left[g_{1}^{2}|U_{1},U_{2}\right].
\end{align*}
The last inequality is due to condition \ref{enu:L2inequal} with $\rho \in (0,1)$. Consequently,
Lemma 8.8 in \citep{Vaart2002} yields $\nu_{1}\left\{ g_{1}h_{1}+g_{2}\right\} ^{2}\gtrsim\nu_{1}\left(g_{1}^{2}\right)+\nu_{1}\left(g_{2}^{2}\right)$.
Under \ref{enu:paraspace} and \ref{enu:identifiability},  by Jensen's inequality, 
$
\nu_{1}\left(g_{1}^{2}\right)\gtrsim\left\Vert \bb-\bb_{0}\right\Vert _{2}^{2}
$.
Therefore, $\nu_{1}\left\{ g_{1}h_{1}+g_{2}\right\} ^{2} \gtrsim d^{2}\left(\bt,\bt_{0}\right)$ and so does 
$\mathbf{M}\left(\bt_{0}\right)-\mathbf{M}\left(\bt\right)$.

Next, we need to derive $\phi_{n}\left(\eta\right)$ such that 
$
E\sup_{d\left(\bt,\bt_{0}\right)<\eta}\left|\mathbb{G}_{n}\left(m_{\bt}-m_{\bt_{0}}\right)\right|\lesssim\phi_{n}\left(\eta\right).
$
Define classes $\mathcal{M}_{\eta}=\left\{ m_{\bt}-m_{\bt_{0}}:d\left(\bt,\bt_{0}\right)<\eta,\bt\in\mathcal{B}\times\mathcal{F}\right\} $
and $\mathcal{F}_{\eta}=\{ \bL\in\mathcal{F}:\Vert \Delta\bL-\Delta\bL_{0}\Vert _{L_{2}(\mu_{1})}\le\eta\} $.
We start by obtaining the bracketing number of $\mathcal{M}_{\eta}$ in the following lemma whose proof is provided in Appendix~\ref{app:proofs}.
\begin{lem}
	Under the conditions 
	\ref{enu:paraspace}, \ref{enu:xbounded}, \ref{enu:separable}, \ref{enu:finite_obs},
	 \ref{enu:expbound} and \ref{enu:mu2con}, $\log N_{[\,]}\left(\epsilon,\mathcal{M}_{\eta},\left\Vert \cdot\right\Vert _{P,B}\right)\lesssim1/\epsilon$, where the Bernstein norm  $\Vert f \Vert _{P,B}=\{ 2\mathbf{P}(\exp(|f|)-1-|f|)\} ^{1/2}\le\{ \mathbf{P}(\exp(|f|)|f|^{2})\} ^{1/2}$. \label{lem:bracketingnumber}
\end{lem}
It follows from Lemma~\ref{lem:bracketingnumber}, $\tilde{J}_{[\,]}\left(\eta,\mathcal{M}_{\eta},\left\Vert \cdot\right\Vert _{P,B}\right)=\int_{0}^{\eta}\sqrt{1+\log N_{[\,]}\left(\epsilon,\mathcal{M}_{\eta},\left\Vert \cdot\right\Vert _{P,B}\right)}d\epsilon\lesssim\delta^{1/2}$.
The similar arguments to derive the bracketing number of $\mathcal{M}_{\eta}$
can also yield that $\left\Vert f\right\Vert _{P,B}\lesssim\eta$
for every $f\in\mathcal{M}_{\eta}$. By Lemma 3.4.3 of \citep{Vaart1996},
we have $E_{\mathbf{P}}\Vert \mathbb{G}_{n}\Vert _{\mathcal{M}_{\eta}}\lesssim\tilde{J}_{[\,]}(\eta,\mathcal{M}_{\eta},\Vert \cdot\Vert _{P,B})\{ 1+\tilde{J}_{[\,]}(\eta,\mathcal{M}_{\eta},\Vert \cdot\Vert _{P,B})/(\eta^{2}\sqrt{n})\} \lesssim\eta^{1/2}+\eta^{-1}/\sqrt{n}$,
leading to $\phi_{n}\left(\eta\right)=\eta^{1/2}+\eta^{-1}/\sqrt{n}$. 

It is easy to see $\phi_{n}\left(\eta\right)/\eta$ is a decreasing
function.
If we let $r_{n}=n^{1/3}$, $r_{n}^{2}\phi_{n}(1/r_{n})=n^{2/3}(n^{-1/6}+n^{-1/6})=2\sqrt{n}$.
By Theorem 3.2.5 in \citet{Vaart1996}, $n^{1/3}d(\hat{\bt},\bt)=O_{p}\left(1\right)$
which completes the proof.

\subsection{Proof of Theorem~\ref{thm:normality}}

We consider a submodel $\left(\bb+\epsilon h_{1},\int\left(1+\epsilon h_{2}\right)d\bL\right)$
where $\mathcal{H}_{1}=\left\{ h_{1}:h_{1}\in\mathcal{B},\left\Vert h\right\Vert _{2}\le1\right\} $
and $\mathcal{H}_{2}=\{ h_{2}:h_{2}$ is a function with bounded total variation in $\left[0,\tau\right]\} $.
Let $\mathcal{H}=\mathcal{H}_{1}\times\mathcal{H}_{2}$ and define a sequence of maps $S_{n}$ mapping a neighborhood
of $\bt_{0}$, denoted by $\mathcal{U}$, in the parameter space
$\Theta$ into $l^{\infty}\left(\mathcal{H}\right)$
as 
$
	S_{n}(\bt)[h_{1},h_{2}]= \frac{d}{d\epsilon}\frac{1}{n}\ell(\bb+\epsilon h_{1},\int(1+\epsilon h_{2})d\bL)|_{\epsilon=0}	= \mathbb{P}_{n}m_{\bt}^{\prime}[h_{1},h_{2}],
$
where
$m_{\bt}^{\prime}[h_{1}]=m_{1\bt}^{\prime}[h_{1}]+m_{2\bt}^{\prime}[h_{1}]$ with %
$
m_{1\bt}^{\prime}[h_{1}]=\sum_{j=1}^{J}(\Delta N_{\cdot j}/D_{j}^{0}(\bt)-1)D_{j}^{1}(\bt,h_{1}^{\top}\bX))
$
and 
$
m_{2\bt}^{\prime}[h_{2}]=\sum_{j=1}^{J}(\Delta N_{\cdot j}/D_{j}^{0}(\bt)-1)D_{j}^{1}(\bt,h_{2}).
$
Set $S(\bt)[h_{1},h_{2}]\allowdisplaybreaks=\mathbf{P}m_{\bt}^{\prime}\left[h_{1},h_{2}\right]$.

Inspired by \citet{Liu2021}, we need to verify the following conditions
to derive the asymptotic normality of $\hat{\bb}$: 
\begin{enumerate}[label=(a\arabic*)]
	\item $\sqrt{n}(S_{n}-S)(\hat{\bt})[h_{1},h_{2}]-\sqrt{n}(S_{n}-S)(\bt_{0})[h_{1},h_{2}]=o_{p}(1)$. 
	\item $S(\bt_{0})[h_{1},h_{2}]=0$ and $S_{n}(\hat{\bt})[h_{1},h_{2}]=o_{p}(n^{-1/2})$. 
	\item $\sqrt{n}(S_{n}-S)(\bt_{0})[h_{1},h_{2}]$
	converges in distribution to a tight Gaussian process on $l^{\infty}(\mathcal{H})$. 
	\item $S(\bt)[h_{1},h_{2}]$ is Fr\'{e}chet-differentiable
	at $\bt_{0}$ with a continuous derivative $\dot{S}(\bt_{0})(\bt-\bt_{0})[h_{1},h_{2}]$. 
	\item $\sqrt{n}\{ S(\hat{\bt})[h_{1},h_{2}]-S(\bt_{0})[h_{1},h_{2}]-\dot{S}(\bt_{0})(\hat{\bt}-\bt_{0})[h_{1},h_{2}]\} =o_{p}(1)$. 
\end{enumerate}
The following proposition provides a general form of the asymptotic
distribution of $\hat{\bt}$. 
\begin{prop}\label{prop:genasym}
Under conditions (a1)-(a2) and (a4)-(a5), 
	$
	-\sqrt{n}\dot{S}(\bt_{0})(\hat{\bt}-\bt_{0})[h_{1},h_{2}]=\sqrt{n}(S_{n}-S)(\bt_{0})[h_{1},h_{2}]+o_{p}(1)
	$.
\end{prop}
The proof of Proposition \ref{prop:genasym} is provided in Appendix~\ref{app:proofs}.

To verify (a1), for $\eta_{n}$ such that $d(\hat{\bt},\bt_{0})/\eta_{n}\to0$,
define
$
	\Psi_{n}=\{ \psi(\bt;h_{1},h_{2})=m_{\bt}^{\prime}[h_{1},h_{2}]-m_{\bt_{0}}^{\prime}[h_{1},h_{2}],
	d(\bt,\bt_{0})<\eta_{n},\bt\in\mathcal{B}\times\mathcal{F},h_{1}\times h_{2}\in\mathcal{H}\} $.	
Since the total variation of $h_{2}$ is bounded, by Corollary 2.7.2
in \citet{Vaart1996}, $\log N_{[\,]}\left(\epsilon,\mathcal{H}_{2},L_{2}\left(\mathbf{P}\right)\right)\lesssim1/\epsilon$.
By the similar bracketing entropy argument in the proof of Theorem~\ref{thm:rate}, we can deduce that $\log N_{[\,]}\left(\epsilon,\Psi_{n},L_{2}\left(\mathbf{P}\right)\right)\lesssim1/\epsilon$.
Then, 
$
	J_{[\,]}\left(\delta_{n},\Psi_{n},L_{2}\left(\mathbf{P}\right)\right) =\int_{0}^{\delta_{n}}\sqrt{\log N_{[\,]}\left(\epsilon,\Psi_{n},L_{2}\left(\mathbf{P}\right)\right)}d\epsilon \lesssim\int_{0}^{\delta_{n}}\sqrt{1/\epsilon}d\epsilon\lesssim\delta_{n}^{1/2}
$.
Clearly, $J_{[\,]}\left(\delta_{n},\Psi_{n},L_{2}\left(\mathbf{P}\right)\right)\to0$
for every sequence $\delta_{n}\to0$. Moreover, 
$
\sup_{(h_{1},h_{2})\in\mathcal{H}}\mathbf{P}(\psi(\hat{\bt};h_{1},h_{2})-\psi(\bt_{0};h_{1},h_{2}))^{2}\to0
$,
in probability as $d(\hat{\bt},\bt_{0})\to0$ because
$\psi\left(\bt;h_{1},h_{2}\right)$ is Lipschitz continuous in
$\bt$ under $L_{2}\left(\mathbf{P}\right)$ by \ref{enu:paraspace},
\ref{enu:xbounded}, \ref{enu:separable} and \ref{enu:expbound}.
It is not hard to show that $\Pr\left(\psi\left(\bt_{n};h_{1},h_{2}\right)\in\Psi_{n}\right)$
goes to 1 as $n\to\infty$ by the definition of $\eta_{n}$.
It follows from Theorem 2.2 in \citet{Vaart2007} that 
\[
\sqrt{n}\left|\left(S_{n}-S\right)\left(\hat{\bt}\right)\left[h_{1},h_{2}\right]-\left(S_{n}-S\right)\left(\bt_{0}\right)\left[h_{1},h_{2}\right]\right|\le\sup_{\left(h_{1},h_{2}\right)\in\mathcal{H}}\sqrt{n}\left|\left(\mathbb{P}-\mathbf{P}\right)\psi\left(\hat{\bt};h_{1},h_{2}\right)\right|\to_{p}0,
\]
and hence (a1) follows.

To verify (a2), it is not hard to show $S(\bt_{0})[h_{1},h_{2}]=0$
after some algebra. For $S_{n}(\hat{\bt})[h_{1},h_{2}]$,
let $S_{n}(\hat{\bt})[h_{1},h_{2}]=S_{1n}(\hat{\bt})[h_{1}]+S_{1n}(\hat{\bt})[h_{2}]$
where $S_{kn}\left(\bt\right)\left[h_{k}\right]=m_{k\bt}^{\prime}\left[h_{k}\right]$, for $k=1,2$.
The first part $S_{1n}(\hat{\bt})\left[h_{1}\right]=0$
holds as $S_{1n}(\hat{\bt})\left[h_{1}\right]$ is the
score function for $\bb$. For $S_{2n}\left(\bt\right)\left[h_{2}\right]$,
we consider a submodel $\bt_{\epsilon}=\left(\hat{\bb},\int\left(1+\epsilon h_{2}\right)d\hat{\bL}\right)$
for $h_{2}\in\mathcal{H}_{2}$. By the definition of Riemann--Stieltjes
integral $\int\left(1+\epsilon h_{2}\right)d\hat{\bL}\in\mathcal{F}^{*}$
for $h_{2}\in\mathcal{H}_{2}$ and small enough $\epsilon$. Since
$\hat{\bt}$ maximizes $\mathbf{\mathbb{P}}_{n}m_{\bt}$ over
$\mathcal{B}\times\mathcal{F}^{*}$, we find that $0=\lim_{\epsilon\to0}d\mathbf{\mathbb{P}}_{n}m_{\bt_{\epsilon}}/d\epsilon=S_{2n}(\hat{\bt})[h_{2}]$,
for any $h_{2}\in\mathcal{H}_{2}$. Therefore, $n^{1/2}S_{n}(\hat{\bt})[h_{1},h_{2}]=o_{p}(1)$
because $S_{n}(\hat{\bt})[h_{1},h_{2}]=0$.

For (a3), let 
$
\bar{\Psi}=\{ m_{\bt_{0}}^{\prime}[h_{1},h_{2}],h_{1}\times h_{2}\in\mathcal{H}\}
$.
It is easily seen that $\mathcal{H}_{1}$ and $\mathcal{H}_{2}$ are $\mathbf{P}$-Donsker
classes \citet[][Example 2.10.27]{Vaart1996}. %
It can be shown that $m_{\bt_{0}}^{\prime}\left[h_{1},h_{2}\right]$
is Lipschitz continuous with respect to $h_{1}$ and $h_{2}$ under
conditions \ref{enu:paraspace}, \ref{enu:xbounded}, \ref{enu:separable}
and \ref{enu:finite_obs}. By Theorem 2.10.6 in \citep{Vaart1996},
$\bar{\Psi}$ is a $\mathbf{P}$-Donsker class and we have (a3) by
the definition of the Donsker class.

The Fr\'{e}chet differentiability of $S\left(\bt\right)$ holds obviously
by the continuity of $S\left(\bt\right)$ in $\bt$.
Let the Fr\'{e}chet
derivative of $S\left(\bt\right)$ at $\bt_{0}$ denote by
$\dot{S}\left(\bt_{0}\right)\left(\bt-\bt_{0}\right)=d\left.S\left(\bt_{0}+\epsilon\left(\bt-\bt_{0}\right)\right)/d\epsilon\right|_{\epsilon=0}$ that is a map from the space $\left\{ \bt-\bt_{0}:\bt\in\mathcal{U}\right\} $
to $l^{\infty}\left(\mathcal{H}\right)$.
Then, 
$
	 \dot{S}\left(\bt_{0}\right)\left(\bt-\bt_{0}\right)\left[h_{1},h_{2}\right]=  \left(\bb-\bb_{0}\right)^{\top}Q_{1}\left(h_{1},h_{2}\right)+\int_{0}^{\tau}Q_{2}\left(h_{1},h_{2},t\right)d\left(\bL-\bL_{0}\right)
$,
where $Q_{1}(h_{1},h_{2})=-\mathbf{P}\sum_{j=1}^{J}D_{j}^{1}(\bt_{0},h_{1}^{\top}\bX+h_{2})D_{j}^{1}(\bt_{0},\bX)/D_{j}^{0}(\bt_{0})$ and $Q_{2}(h_{1},h_{2},t)=-\mathbf{P}\sum_{j=1}^{J}D_{j}^{1}(\bt_{0},h_{1}^{\top}\bX+h_{2})/D_{j}^{0}(\bt_{0}) I(T_{\cdot(j-1)}<t<T_{\cdot j})\exp(\bb_{0}^{\top}\bX(t))$.

Last for (a5), by using the mean value theorem twice, we can show
under conditions \ref{enu:paraspace}, \ref{enu:xbounded}, \ref{enu:separable},
\ref{enu:finite_obs} and \ref{enu:expbound}, 
$|S(\hat{\bt})[h_{1},h_{2}]-S(\bt_{0})[h_{1},h_{2}]-\dot{S}(\bt_{0})(\hat{\bt}-\bt_{0})[h_{1},h_{2}]|\lesssim O(d^{2}(\hat{\bt},\bt))$.
It is seen that $\sqrt{n}d^{2}(\hat{\bt},\bt)=o_{p}(n^{1/2-2/3})=o_{p}\left(1\right)$
and thus we verify (a5).

After verifying (a1)-(a5), we next show that there
exists a unique $h^{*}_{2}$ such that 
$
Q_{2}(h_{1},h^{*}_{2},t)=0
$ for any $h_{1}\in\mathcal{H}_1$.
Define a bilinear mapping $\varphi:\mathcal{H}_{2}\times\mathcal{H}_{2}\to\mathbb{R}$,
such that, for $v_{1},v_{2}\in\mathcal{H}_{2}$, $\varphi\left(v_{1},v_{2}\right)=\mathbf{P}\sum_{j=1}^{J}D_{j}^{1}(\bt_{0},v_{1})D_{j}^{1}(\bt_{0},v_{2})/D_{j}^{0}(\bt_{0})$
and a linear mapping $b:\mathcal{H}_{2}\to\mathbb{R}$, $b(v;h_{1})=\varphi(-h_{1}^{\top}\bX(t),v)$.
Obviously, $\varphi\left(v_{1},v_{2}\right)$ is linear in $v_{1}$ and
$v_{2}$,  $\varphi\left(v,v\right)\ge0$, and $\varphi\left(v_{1},v_{2}\right)=\varphi\left(v_{2},v_{1}\right)$.
Furthermore, if $\varphi\left(v,v\right)=0$
for $v\in\mathcal{H}_{2}$, then 
$\mathbf{P}\sum_{j=1}^{J}\{D_{j}^{1}(\bt_{0},v)\}^{2}/D_{j}^{0}(\bt_{0})$.
This suggest that $v\left(t\right)=0$ a.e. on $\nu_{1}$
because $\exp(\bb_{0}^{\top}\bX\left(t\right))$ and $d\bL_{0}\left(t\right)$
are positive on $\nu_{1}$ by condition \ref{enu:paraspace}. Hence,
$\varphi\left(v_{1},v_{2}\right)$ is an inner product on $\mathcal{H}_{2}$
and we can define a norm on $\mathcal{H}_{2}$ such that $\left\Vert v\right\Vert _{\mathcal{H}_{2}}^{2}=\varphi\left(v,v\right)$
for $v\in\mathcal{H}_{2}$. As a result, the mapping $\varphi$ is continuous
and bounded. In addition, $b\left(v;h_{1}\right)$ is obviously a
continuous linear operator for any $h_{1}$. Thus, by the F. Ries
representation theorem \citep[Theorem 4.6-1]{Ciarlet2013}, given $h_{1}$, there
exists a unique $h_{2}^{*}\in\mathcal{H}_{2}$ such that $\varphi\left(h_{2}^{*},h_{2}\right)=b\left(h_{2};h_{1}\right)$
for any $h_{2}\in\mathcal{H}_{2}$. Recall that $d\bL-d\bL_{0}$
can expressed as $\epsilon h_{2}d\bL_{0}$ for $h_{2}\in\mathcal{H}_{2}$
and some $\epsilon>0$. Then, 
$ 	\int_{0}^{\tau}Q_{2}\left(h_{1},h_{2}^{*},t\right)d\left(\bL-\bL_{0}\right)  %
=\epsilon\left\{ b\left(h_{2};h_{1}\right)-\varphi\left(h_{2}^{*},h_{2}\right)\right\} 
=0$
for any $\bL\in\mathcal{F}$.

We now give the definition of $\bar{\bx}$.
Solving $h_{2}^{*}$ from $\varphi\left(h_{2}^{*},h_{2}\right)=b\left(h_{2};h_{1}\right)$
for a given $h_{1}$ can be seen as a unique operator $H:\mathcal{H}_{1}\to\mathcal{H}_{2}$
that is linear and continuous in $h_{1}$ by the F. Ries representation
theorem. Hence, we can write $h^{*}_{2}\left(t\right)=H h_1=-h_{1}^{\top}\bar{\bx}\left(t\right)$
where $\bar{\bx}\left(t\right)$ is a $p$-dimensional 
function on $\left[0,\tau\right]$ whose $l$th element is the unique
solution of $-h_{1}^{\top}\varphi(\bar{\bx}\left(t\right),h_{2})=b\left(h_{2};-e_{l}\right)$ for any $h_2$.
Recall that $\mathcal{H}_{3}=\{ h_{3}(t)\in\mathbf{R}^{p},t\in\left[0,\tau\right]:$ each element of $h_{3}(t)$ is of bounded variation over $\left[0,\tau\right]$$\}$. 
Since $h_1$ is arbitrary, 
we can then define $\bar{\bx}$ as the unique solution of
\begin{equation}
\mathbf{P}\sum_{j=1}^{J}D_{j}^{1}(\bt_{0},\bX-h_{3})D_{j}^{1}(\bt_{0},h_{2})/D_{j}^{0}(\bt_{0})=0, \label{eq:xbardefeq}
\end{equation}
over $h_3\in\mathcal{H}_3$ for any $h_2 \in \mathcal{H}_2$.

Next, we derive $\bO$ from $Q_{1}$. Plugging $h^{*}_{2}\left(t\right)=-h_{1}^{\top}\bar{\bx}\left(t\right)$
in $Q_{1}$, we have $Q_{1}\left(h_{1},h^{*}_{2}\right)=-h_{1}^{\top}\bO$ such that
 $\bO=-\mathbf{P}\sum_{j=1}^{J}D_{j}(\bt_{0})D_{j}^{1}(\bt_{0},\bX)^{\top}/D_{j}^{0}(\bt_{0})$ where we define $
 D_{j}(\bt)=D_{j}^{1}(\bt,\bX-\bar{\bx})
 $.
Moreover, we can choose $h_{2}$ as $H h_1^\prime=-h_{1}^{\prime\top}\bar{\bx}\left(t\right)$ for any $h_{1}^{\prime}\in\mathcal{H}_{1}$ and $h_3=\bar{\bx}$.
It follows from \eqref{eq:xbardefeq} that $\mathbf{P}\sum_{j=1}^{J}D_{j}(\bt_{0})D_{j}^{1}(\bt_{0},\bar{\bx})^{\top}/D_{j}^{0}(\bt_{0})=0$.
Adding this matrix to $\bO$, after some algebra, we obtain that $\bO=-E[\sum_{j=1}^{J}D_{j}(\bt_{0})^{\otimes2}/D_{j}^{0}(\bt_{0})]$, coinciding with the form in Theorem~\ref{thm:normality}.
By condition \ref{enu:identifiability}, $-\bO$ is positive definite
and nonsingular, i.e., $\bO^{-1}$ exists.

Then, by condition (a3) and Proposition~\ref{prop:genasym}, for any $h_1$,
\begin{align}
	\sqrt{n}h_{1}^{\top}\bO\left(\hat{\bb}-\bb_{0}\right) & =-\sqrt{n}\left(S_{n}-S\right)\left(\bt_{0}\right)\left[h_{1},h_{2}^{*}\right]+o_{p}\left(1\right)\to_{d}N\left(0,\sigma^{2}\right),\label{eq:asymptoticlinear}
\end{align}
where 
$
\sigma^{2} =E\left[m_{\bt_{0}}^{\prime}[h_{1},h_{2}^{*}]^{2}\right] =h_{1}^{\top}\bS h_{1}
$
and
\begin{equation}
	\bS=E\left[\left\{ \sum_{j=1}^{J}\left\{ \Delta N_{\cdot j}-D_{j}^{0}(\bt_{0})\right\} \frac{D_{j}(\bt_{0})}{D_{j}^{0}(\bt_{0})}\right\} ^{\otimes2}\right]
	=E\left[\sum_{j_{1}=1}^{J}\sum_{j_{2}=1}^{J}D_{j_{1}j_{2}}^{02}\frac{D_{j_{1}}(\bt_{0})}{D_{j_{1}}^{0}(\bt_{0})}\frac{D_{j_{2}}(\bt_{0})}{D_{j_{2}}^{0}(\bt_{0})}\right]
	, \label{eq:Sigmadef}
\end{equation}
where $D_{j_{1}j_{2}}^{02}=\mathrm{Cov}[\Delta N_{\cdot j_{1}}\Delta N_{\cdot j_{2}}|T_{\cdot j_{1}},T_{\cdot(j_{1}-1)},T_{\cdot j_{2}},T_{\cdot(j_{2}-1)},\bX,J]$.
Since $h_{1}$ is arbitrary, by the Cram\'{e}r-Wold theorem, $\sqrt{n}\bO(\hat{\bb}-\bb_{0})\to_{d}N\left(0,\bS\right)$.
When the Poisson assumption holds, 
$D_{j_{1}j_{2}}^{02}=D_{j}^{0}(\bt_{0})$ if $j_{1}=j_{2}=j$ and zero otherwise. Hence, after some algebra, $\bS=-\bO$ under the Poisson assumption.

Furthermore, we show that $\hat{\bb}$ achieves the asymptotic efficiency
if the Poisson model holds. Under the Poisson assumption, the log-likelihood
function $\mathbb{P}_{n}m_{\bt}$ would be the ``true'' log-likelihood
function and we can take advantage of the existing conclusions regarding
semiparametric efficiency \citep[see,][]{Vaart2000,Kosorok2008}.
Denote the derivative of $m_{\bt}$ with respect
to $\bb$ by 
$
\dot{m}_{\bt}[h_{1}]=\sum_{j=1}^{J}(\Delta N_{\cdot j}/D_{j}^{0}(\bt)-1)h_{1}^{\top}D_{j}^{1}(\bt,\bX).
$
Denote
the score operator for $\bL$ and its adjoint operator by $B_{\bb,\bL}$
and $B_{\bb,\bL}^{*}$, respectively \citep[][Chapter 25]{Vaart2000}.
Then, with the submodel $d\bL_{\epsilon}=\left(1+\epsilon h_{2}\right)d\bL_{0}$,
Theorem 1 in \citet{Mao2018} yields that
$
B_{\bb,\bL}^{*}\dot{m}_{\bt}[h_{1}](\cdot)=-\sum_{j=1}^{J}\int_{T_{\cdot(j-1)}}^{T_{\cdot j}}E_{\bt}[\exp(\bb^{\top}\bX(u))h_{1}^{\top}\bX(u)\exp(\bb^{\top}\bX(\cdot))/D_{j}^{0}(\bt)]d\bL(u),
$
and 
$
B_{\bb,\bL}^{*}B_{\bb,\bL}h_{2}(\cdot)=\sum_{j=1}^{J}\int_{T_{\cdot(j-1)}}^{T_{\cdot j}}E_{\bt}[\exp(\bb^{\top}\bX(u))\exp(\bb^{\top}\bX(\cdot))/D_{j}^{0}(\bt)]h_{2}(u)d\bL(u)
$.
It is not hard to show that, for any $h_{1}\in\mathcal{H}_{1}$ there
exists $h_{2}^{*}$ such that $\int_{0}^{\tau}Q_{2}\left(h_{1},h_{2}^{*},t\right)d\left(\bL-\bL_{0}\right)=0$
for any $\bL\in\mathcal{F}$ implies that the normal equation
$B_{\bb,\bL}^{*}\dot{m}_{\bt}\left[h_{1}\right]\left(\cdot\right)=B_{\bb,\bL}^{*}B_{\bb,\bL}h_{2}^{*}\left(\cdot\right)$
holds.
It follows from $h^{*}_{2}=h_1^{\top}\bar{\bx}$ that $\bar{\bx}$  is a least favorable direction vector.
Clearly, the efficient
score at $\bt_0$ is then
\begin{equation}
	\tilde{m}_{\bt_{0}}=\sum_{j=1}^{J}(\Delta N_{\cdot j}-D_{j}^{0}(\bt_{0}))D_{j}(\bt_{0})/D_{j}^{0}(\bt_{0}).\label{eq:efficientscore}
\end{equation}

By \eqref{eq:asymptoticlinear} and some algebra, 
$
\sqrt{n}(\hat{\bb}-\bb)=\tilde{I}_{\bb_{0},\bL_{0}}^{-1}\mathbb{G}_{n}\tilde{m}_{\bt_{0}}+o_{p}(1)
$
where $\tilde{I}_{\bb_{0},\bL_{0}}$ is the efficient information
matrix for $\bb$ and $\tilde{I}_{\bb_{0},\bL_{0}}=-\mathbf{P}(\partial\tilde{m}_{\bt}/\partial\bb|_{\bb=\bb_{0},\bL=\bL_{0}})=-\bO$.
Then, $\sqrt{n}(\hat{\bb}-\bb)$ is asymptotic linear
in the efficient influence function $\tilde{I}_{\bb_{0},\bL_{0}}^{-1}\tilde{m}_{\bt_{0}}$.
Therefore, $\hat{\bb}$ is  asymptotic efficient under Poisson assumption \citep[][Lemma 25.23]{Vaart2000}. 

\subsection{The variance estimation derivation and proof of Theorem~\ref{thm:varest}}\label{sec:proofvarest}

In this section, we derive the variance estimators proposed in Section \ref{sec:varianceestimation}. We prove their consistency in Theorem~\ref{thm:varest} and provide the proof of Lemma~\ref{lem:xbarconsist}.

\subsubsection{Proof of Theorem~\ref{thm:varest}}

In the proof of Theorem
\ref{thm:normality}, \eqref{eq:Sigmadef}, \eqref{eq:efficientscore} and some algebra show that
$\mathbf{P}\left.\partial\tilde{m}_{\bt}/\partial\bb\right|_{\bt=\bt_{0}}=\bO$ 
that
 $\bS=\mathbf{P}\left(\tilde{m}_{\bt_{0}}\right)^{\otimes2}$
despite the validity of the Poisson assumption.
In other words,
both $\bO$ and $\bS$ can be derived from the efficient score function $\tilde{m}_{\bt_{0}}$ defined in \eqref{eq:efficientscore}  regardless of the Poisson model. 
By this observation, the main idea of the proof is to derive the alternative forms of $\bO$ and $\bS$ expressed by $t_{k}$ and $\lambda_{k}$ by connecting $\tilde{m}_{\bt}$ to the EM algorithm.
Facilitated by the alternative forms, we then show the consistency results in Theorem~\ref{thm:varest}.

We first show the consistency of $\hat{\bS}$.
Recall that $\mathcal{H}_{3}=\{ h_{3}(t)\in\mathbf{R}^{p},t\in\left[0,\tau\right]:$ each element of $h_{3}(t)$ is of bounded variation over $\left[0,\tau\right]$$\}$.
Define $\tilde{m}^{EM}[\bt;h_{3}]=\sum_{k=1}^{K}\Delta_{\cdot k}(\hat{E}[W_{\cdot k}|\bb,\bl]-\exp(\bb^{\top}\bX(t_{k}))\lambda_{k})(\bX(t_{k})-h_{3}(t_{k}))$.
Clearly, $\hat{\bS}=\mathbb{P}_{n}(\tilde{m}^{EM}[\hat{\bt};\hat{\bar{\bx}}])^{\otimes2}$.
We next establish the connection between $\tilde{m}^{EM}\left[\bt;h_{3}\right]$ and the efficient score function $\tilde{m}_{\bt}$.
Using the arguments to discretizing $\Lambda(t)$ by $\bl$ on $\{ t_{k}\} _{k=1}^{K}$, we have
\begin{align*}
	& \sum_{j=1}^{J}\left(\Delta N_{\cdot j}/D_{j}^{0}(\bt)-1\right)D_{j}^{1}(\bt,\bX-h_{3})\\
	= & \sum_{k=1}^{K}\Delta_{\cdot k}\left(\frac{\Delta_{\cdot k}\exp\left(\bb^{\top}\bX\left(t_{k}\right)\right)\lambda_{k}\Delta N_{\cdot j_{\cdot k}}}{\sum_{l\in A_{\cdot k}}\Delta_{\cdot k}\exp\left(\bb^{\top}\bX\left(t_{l}\right)\right)\lambda_{l}}-\exp\left(\bb^{\top}\bX\left(t_{k}\right)\right)\lambda_{k}\right)\left(\bX\left(t_{k}\right)-h_{3}\left(t_{k}\right)\right)\\
	\overset{\eqref{eq:EWdef}}{=} & \sum_{k=1}^{K}\Delta_{\cdot k}\left(\hat{E}\left[W_{\cdot k}|\bb,\bl\right]-\exp\left(\bb^{\top}\bX\left(t_{k}\right)\right)\lambda_{k}\right)\left(\bX\left(t_{k}\right)-h_{3}\left(t_{k}\right)\right)=\tilde{m}^{EM}\left[\bt;h_{3}\right].
\end{align*}
By setting $h_3=\bar{\bx}$, we have $\tilde{m}^{EM}\left[\bt;\bar{\bx}\right]=\tilde{m}_{\bt}$. Consequently, we obtain $\bS=\mathbf{P}\left(\tilde{m}^{EM}\left[\bt_{0};\bar{\bx}\right]\right)^{\otimes2}$. 
Since each element of $h_3$ belongs to the bounded total variation space $\mathcal{H}_3$, we can use bracketing entropy arguments similar to those in the proof of Theorem~\ref{thm:rate}-\ref{thm:normality} to show that the set 
$ 
\Phi^{EM}_{\eta}=\{ \tilde{m}^{EM}\left[\bt;h_{3}\right], d(\bt,\bt_{0})<\eta,\bt\in\mathcal{B}\times\mathcal{F},h_{3} \in \mathcal{H}_3\}$,
is a $\mathbf{P}$-Donsker class and Glivenko-Cantelli class under
conditions \ref{enu:paraspace}, \ref{enu:xbounded}, \ref{enu:separable},
\ref{enu:finite_obs} and \ref{enu:expbound}.
By Lemma~\ref{lem:xbarconsist} and Theorem~\ref{thm:consistency}, $\hat{\bar{\bx}}$ converges to $\bar{\bx}$ in probability and  $\hat{\bt}$ converges to $\bt_{0}$. It then follows from the continuity of $\tilde{m}^{EM}[\bt;h_3]$ in $\bt$ and $h_3$ that $\hat{\bS}$ converges to $\bS$ in probability.

Next, we establish the consistency of $\hat{\bO}$.
Likewise, we
need to derive the form of $\bO$ in terms of $t_{k}$ and $\bl_{k}$
first. However, it is not straightforward due to the mathematical intractability of the explicit
form of $d\hat{\lambda}_{k}\left(\bb\right)/d\bb$. From the proof
of Theorem~\ref{thm:normality} on page \pageref{eq:asymptoticlinear}, $-\bO=\bS$ when the Poisson model holds and the
form of $\bO$ stays the same no matter if the Poisson assumption
holds. Thus, we can find the form of $\bO$ by explicitly calculating
$\bS$ under the Poisson assumption. Recalling the form of $\bS$, under the Poisson model,
some algebra yields that 
\[
\bO=-E\left[\sum_{j=1}^{J}\frac{\left\{ \sum_{T_{\cdot\left(j-1\right)}<t_{k}\le T_{\cdot j}}\exp\left(\bb_{0}^{\top}\bX\left(t_{k}\right)\right)\lambda_{0k}\left(\bX\left(t_{k}\right)-\bar{\bx}\left(t_{k}\right)\right)\right\} ^{\otimes2}}{\sum_{T_{\cdot\left(j-1\right)}<t_{k}\le T_{\cdot j}}\exp\left(\bb_{0}^{\top}\bX\left(t_{k}\right)\right)\lambda_{0k}}\right].
\]
Similarly to the arguments to prove the consistency of $\hat{\bS}$, we can show $\hat{\bO}$,
the empirical counterpart of $\bO$, converges to $\bO$
in probability. 

Last, we prove that  $\hat{\bV}$ converges to $\bS$ in probability and is asymptotically equal to $\hat{\bS}$. 
We begin from expressing $\partial\mathrm{p}\ell_{i}\left(\bb|\mathcal{O}\right)/\partial\bb$ in the context of the EM algorithm using equation (5) in \citet{Oakes1999}.
We can write $\mathrm{p}\ell_{i}\left(\bb|\mathcal{O}\right)=\ell_{i}(\bb,\hat{\bL}_{\bb}|\mathcal{O})$.
Equation (5) in \citet{Oakes1999} shows that the derivatives of  $\ell_{i}\left(\bb,\bL|\mathcal{O}\right)$ with respect to (w.r.t.) $\bt$ is the derivative of $E_{C|O,\bt^{\prime}}\left[\ell_{i}\left(\bb,\bl|\mathcal{C}\right)\right]$ w.r.t. $\bt$ given that $\bt^{\prime}=\bt$, where  $E_{C|O,\bt}$ denotes the conditional expectation the complete data $\mathcal{C}$ given the observed data $\mathcal{O}$ with parameter $\bt$.
Apply the chain rule and (5) in \citet{Oakes1999} to $\partial\mathrm{p}\ell_{i}\left(\bb|\mathcal{O}\right)/\partial\bb$, it follows from $\hat{\bar{\bx}}_{k}=-\left.d\log(\hat{\lambda}_{k}(\bb))/d\bb\right|_{\bb=\hat{\bb}}$ that $\left.\partial\mathrm{p}\ell_{i}\left(\bb|\mathcal{O}\right)/\partial\bb\right|_{\bb=\hat{\bb}}=\tilde{m}^{EM}[\hat{\bt};\hat{\bar{\bx}}]$.
Since  $\hat{\bS}=\mathbb{P}_{n}(\tilde{m}^{EM}[\hat{\bt};\hat{\bar{\bx}}])^{\otimes2}$, $\hat{\bS}=\frac{1}{n}\sum_{i=1}^{n}\{ \partial\mathrm{p}\ell_{i}(\bb|\mathcal{O})/\partial\bb|_{\bb=\hat{\bb}}\} ^{\otimes2}$.
Finally, by the definition of $\hat{\bV}$ and Taylor's expansion, $\hat{\bV}=\hat{\bS}+O_{p}\left(h_{n}^{2}\right)$.
Given $h_{n}=O(n^{-1/2})$, $\hat{\bV}=\hat{\bS}+O_{p}\left(n^{-1}\right)$,
implying that $\hat{\bV}$ and $\hat{\bS}$ are asymptotically equal
in probability. Thus, $\hat{\bV}$ converges to $\bS$ in probability
due to the convergence of $\hat{\bS}$.

\subsubsection{The proof of Lemma \ref{lem:xbarconsist}}

Define a class $\Phi_{\eta}^{*}=\{ \varrho(\bt,h_{3})=-\sum_{j=1}^{J}\Delta N_{\cdot j}\Vert D_{j}^{1}(\bt,\bX-h_{3})/D_{j}^{0}(\bt)\Vert_{2}^{2},
d(\bt,\bt_{0})<\eta,\bt\in\mathcal{B}\times\mathcal{F},h_{3}\in\mathcal{H}_{3}\}$.
It follows that $\bar{\bx} \in \mathcal{H}_{3}$ since $h_{1}\bar{\bx}(t)\in \mathcal{H}_2$ for any $h_{1}\in\mathcal{H}_1$ from the proof of Theorem~\ref{thm:normality}. 
Given model~\eqref{eq:ratemodel}, we have that
$\mathbf{P}\varrho(\bt_{0},h_{3})=-\mathbf{P}\sum_{j=1}^{J}\Vert D_{j}^{1}(\bt_{0},\bX-h_{3})\Vert _{2}^{2}/D_{j}^{0}(\bt_{0})$
and  $\mathbf{P}\varrho\left(\bt_{0},\bar{\bx}\right)=-\mathbf{P}\sum_{j=1}^{J}\left\Vert D_{j}(\bt_{0})\right\Vert _{2}^{2}/D_{j}^{0}(\bt_{0})$.
We next verify three conditions of Theorem 2.12 (i) in \citet{Kosorok2008} and then prove the convergence in probability.

\underline{\textbf{Step 1.}} First, we establish that $\liminf_{h_{3}\to\bar{\bx}}\mathbf{P}\varrho\left(\bt_{0},h_{3}\right)\ge\mathbf{P}\varrho\left(\bt_{0},\bar{\bx}\right)$ implies $\tilde{d}\left(\bar{\bx},h_{3}\right) \to 0$ for $h_3\in\mathcal{H}_3$.
It easily follows that for any $h_{3}\in\mathcal{H}_{3}$, $\mathbf{P}\varrho(\bt_{0},\bar{\bx})-\mathbf{P}\varrho(\bt_{0},h_{3})=\mathbf{P}\sum_{j=1}^{J}\{ D_{j}^{1}(\bt_
{0},\bX-h_{3})+D_{j}(\bt_{0})\} ^{\top}D_{j}^{1}(\bt_{0},\bar{\bx}-h_{3})/D_{j}^{0}(\bt_{0})$.
Since each element of $\bar{\bx}(t)-h_{3}(t)$ belongs to $\mathcal{H}_2$ for any $h_3\in\mathcal{H}_3$, it follows from the definition of $\bar{\bx}$ based on \eqref{eq:xbardefeq} that $\mathbf{P}\sum_{j=1}^{J}D_{j}(\bt_{0})^{\top}D_{j}^{1}(\bt_{0},\bar{\bx}-h_{3})/D_{j}^{0}(\bt_{0})=0$.
Subtracting the last equation from $\mathbf{P}\varrho(\bt_{0},\bar{\bx})-\mathbf{P}\varrho(\bt_{0},h_{3})$ twice, we  have 
$\mathbf{P}\varrho(\bt_{0},\bar{\bx})-\mathbf{P}\varrho(\bt_{0},h_{3})=\mathbf{P}\sum_{j=1}^{J}\Vert D_{j}^{1}(\bt_{0},\bar{\bx}-h_{3})\Vert _{2}^{2}/D_{j}^{0}(\bt_{0})$.
It then follows from \ref{enu:paraspace} and \ref{enu:xbounded} that $\mathbf{P}\varrho\left(\bt_{0},\bar{\bx}\right)-\mathbf{P}\varrho\left(\bt_{0},h_{3}\right)\gtrsim \tilde{d}\left(\bar{\bx},h_{3}\right)^{2}$.
Taking the limsup on both sides, we have $\liminf_{h_{3}\to\bar{\bx}}\mathbf{P}\varrho\left(\bt_{0},h_{3}\right)\ge\mathbf{P}\varrho\left(\bt_{0},\bar{\bx}\right)$ implies $\limsup_{h_{3}\to\bar{\bx}}\tilde{d}\left(\bar{\bx},h_{3}\right)\le 0$.
However, $\tilde{d}\left(\bar{\bx},h_{3}\right)\ge 0$ by its definition and thus $\liminf_{h_{3}\to\bt_{0}}\mathbf{P}\varrho\left(\bt_{0},h_{3}\right)\ge\mathbf{P}\varrho\left(\bt_{0},\bar{\bx}\right)$ implies $\tilde{d}\left(\bar{\bx},h_{3}\right) \to 0$.

\underline{\textbf{Step 2.}} We next prove that $\mathbb{P}_{n}\varrho(\hat{\bt},\hat{\bar{\bx}})\ge\sup_{h_{3}\in\mathcal{H}_{3}}\mathbb{P}_{n}\varrho(\hat{\bt},h_{3})-o_{p}(1)$.
It is not hard to show that $\mathbb{P}_{n}\varrho(\hat{\bt},h_{3})$ is concave in $h_3$ in terms of the G\^ateaux-derivative. 
Then, there exists $h^{*}_3$ that maximizes $\mathbb{P}_{n}\varrho(\hat{\bt},h_{3})$ over $\mathcal{H}_{3}$.
By the concavity of $\mathbb{P}_{n}\varrho(\hat{\bt},h_{3})$, for $\epsilon \in [0,1]$,
$
\{ \mathbb{P}_{n}\varrho(\hat{\bt},\hat{\bar{\bx}}+\epsilon(h_{3}^{*}-\hat{\bar{\bx}}))-\mathbb{P}_{n}\varrho(\hat{\bt},\hat{\bar{\bx}})\} /\epsilon\ge\mathbb{P}_{n}\varrho(\hat{\bt},h_{3}^{*})-\mathbb{P}_{n}\varrho(\hat{\bt},\hat{\bar{\bx}})
$.
Furthermore, discretizing $\Lambda(t)$ in $\mathbb{P}_{n}\varrho(\hat{\bt},h_{3})$ by $\bl$ on $\{ t_{k}\} _{k=1}^{K}$ and  letting $\epsilon\to0$
we obtain that 
\begin{multline}
	\frac{2}{n}\sum_{i=1}^{n}\sum_{k=1}^{K}\left\{ \frac{\sum_{l\in A_{ik}}\Delta_{il}\exp\left(\hat{\bb}^{\top}\bX_{i}\left(t_{l}\right)\right)\hat{\lambda}_{l}(\hat{\bb})\left(\bX_{i}\left(t_{l}\right)-\hat{\bar{\bx}}(t_{l})\right)}{\sum_{l\in A_{ik}}\Delta_{il}\exp\left(\hat{\bb}^{\top}\bX_{i}\left(t_{l}\right)\right)\hat{\lambda}_{l}(\hat{\bb})}\right\}^{\top} \\
	\times\left(h_{3}^{*}(t_{k})-\hat{\bar{\bx}}(t_{k})\right)\hat{E}\left[W_{ik}|\hat{\bb},\hat{\bl}(\hat{\bb})\right]\ge\mathbb{P}_{n}\varrho\left(\hat{\bt},h_{3}^{*}\right)-\mathbb{P}_{n}\varrho\left(\hat{\bt},\hat{\bar{\bx}}\right). \label{ref:up_rho}
\end{multline}
We shall prove that the left-hand side of \eqref{ref:up_rho} is $o_p(1)$ by using the properties of the profile log-likelihood function. 

The iterative relationship in \eqref{eq:lambdaupdate}  implies that, for any $\bb$ and $h_{2k} \in \mathcal{H}_2$,
\begin{equation}
	\sum_{i=1}^{n}\sum_{k=1}^{K}h_{2k}\Delta_{ik}\left\{ \hat{E}\left[W_{ik}|\bb,\hat{\bl}(\bb)\right]-\exp\left(\bb^{\top}\bX_{i}\left(t_{k}\right)\right)\hat{\lambda}_{k}(\bb)\right\}=0. \label{eq:Pndr2}
\end{equation}
Thus, recalling the definition of $\hat{E}\left[W_{ik}|\bb,\hat{\bl}(\bb)\right]$ in \eqref{eq:EWdef} and taking differentiation of \eqref{eq:Pndr2} w.r.t. $\bb$, 
after some algebra, we obtain that, for any $\bb$ and $h_{2k} \in \mathcal{H}_2$,
\begin{multline}
\frac{2}{n}\sum_{i=1}^{n}\sum_{k=1}^{K}\Delta_{ik}\left\{ \bX_{i}\left(t_{k}\right)-\hat{\bar{\bx}}(t_{k})\right\} \left\{ \hat{E}\left[W_{ik}|\hat{\bb},\hat{\bl}(\hat{\bb})\right]-\exp\left(\hat{\bb}^{\top}\bX_{i}\left(t_{k}\right)\right)\hat{\lambda}_{k}(\hat{\bb})\right\} h_{2k}\\
	=\frac{2}{n}\sum_{i=1}^{n}\sum_{k=1}^{K}h_{2k}\Delta_{ik}\frac{\sum_{l\in A_{ik}}\Delta_{il}\exp\left(\bb_{0}^{\top}\bX_{i}\left(t_{l}\right)\right)\hat{\lambda}_{l}(\hat{\bb})\left(\bX_{i}\left(t_{l}\right)-\hat{\bar{\bx}}(t_{k})\right)}{\sum_{l\in A_{ik}}\Delta_{il}\exp\left(\hat{\bb}^{\top}\bX_{i}\left(t_{l}\right)\right)\hat{\lambda}_{l}(\hat{\bb})} \hat{E}\left[W_{ik}|\hat{\bb},\hat{\bl}(\hat{\bb})\right].\label{eq:connectofderive}
\end{multline}
Since 
$
E\left[\hat{E}\left[W_{ik}|\bb_{0},\Lambda_{0}\right]|\bX_{i},\left\{ t_{k}\right\} _{k=1}^{K},\Delta_{ik}\right]=\exp\left(\bb_{0}^{\top}\bX_{i}\left(t_{k}\right)\right)\lambda_{0k}
$ under model \eqref{eq:ratemodel},
we have the LHS of \eqref{eq:connectofderive}
converges to zero in probability for any $h_2\in\mathcal{H}_2$ and $h_3\in\mathcal{H}_3$ by the law of large number and the continuous mapping theorem.
In other words, the RHS of \eqref{eq:connectofderive} is $o_p(1)$ for any $h_2\in\mathcal{H}_2$.
It also easily follows that LHS of \eqref{ref:up_rho} coincides with the RHS of \eqref{eq:connectofderive} by letting $h_2$ be the elements of $h_{3}^{*}(t_{k})-\hat{\bar{\bx}}(t_{k})$.
Consequently, $\mathbb{P}_{n}\varrho\left(\hat{\bt},\hat{\bar{\bx}}\right)\ge \mathbb{P}_{n}\varrho\left(\hat{\bt},h_{3}^{*}\right)-o_p(1)$.

\underline{\textbf{Step 3.}} Given that $h_3 \in \mathcal{H}_3$, using similar arguments in the proof of Theorem~\ref{thm:normality}, we can show that $\Phi_{\eta}^{*}$ is $\mathbf{P}$-Donsker and hence Glivenko-Cantelli.

In sum, the conclusions in Steps 1-3 verify all the conditions of conclusion (i) of Theorem 2.12 in \citet{Kosorok2008}.
It then follows from Theorem 2.12 in \citet{Kosorok2008}, $\tilde{d}\left(\hat{\bar{\bx}},\bar{\bx}\right) \to 0$ in probability as $n\to\infty$.

Last, we prove the second part of Lemma~\ref{lem:xbarconsist}.
Given Condtion~\ref{enu:Xconst}, there exists  $\bX_{\cdot j}$ such that $\bX(t)=\bX_{\cdot j}$ for $t \in (T_{\cdot(j-1)},T_{\cdot j}]$.
Then, for any $h_{2} \in \mathcal{H}_{2}$, \eqref{eq:xbardefeq} reduces to 
\begin{align}
	0=& \mathbf{P}\sum_{j=1}^{J}\int_{T_{\cdot\left(j-1\right)}}^{T_{\cdot j}}\exp\left(\bb_{0}^{\top}\bX\left(t\right)\right)\left\{ \bX(t)-h_{3}(t)\right\} d\bL_{0}\left(t\right)\frac{\exp\left(\bb_{0}^{\top}\bX_{\cdot j}\right)\int_{T_{\cdot\left(j-1\right)}}^{T_{\cdot j}}h_{2}(t)d\bL_{0}\left(t\right)}{\exp\left(\bb_{0}^{\top}\bX_{\cdot j}\right)\int_{T_{\cdot\left(j-1\right)}}^{T_{\cdot j}}d\bL_{0}\left(t\right)}\nonumber \\
	= & E\left[\sum_{j=1}^{J}\tilde{h}_{2j}\int_{T_{\cdot\left(j-1\right)}}^{T_{\cdot j}}E\left[\Delta(t)\exp\left(\bb_{0}^{\top}\bX\left(t\right)\right)\left\{ \bX(t)-h_{3}(t)\right\} |T_{\cdot(j-1)},T_{\cdot j},J\right]d\bL_{0}\left(t\right)\right],\label{eq:constantcase}
\end{align}
where $\tilde{h}_{2j}=\int_{T_{\cdot\left(j-1\right)}}^{T_{\cdot j}}h_2(t)d\bL_{0}\left(t\right)/\int_{T_{\cdot\left(j-1\right)}}^{T_{\cdot j}}d\bL_{0}\left(t\right)$.
It follows that $h_{3}(t)=s^{\left(1\right)}(t,\bb_{0})/s^{\left(0\right)}(t,\bb_{0})$.
implies \eqref{eq:constantcase} becomes zero for any $\Lambda \in \mathcal{F}$.
By the uniqueness of $\bar{\bx}$, $\bar{\bx}$ must be $s^{\left(1\right)}(t,\bb_{0})/s^{\left(0\right)}(t,\bb_{0})$ under Condition~\ref{enu:Xconst}.

Under conditions \ref{enu:paraspace} and \ref{enu:xbounded}, it is not
hard to show that $\{ \bX\left(t\right)^{\otimes u}\Delta\left(t\right)\exp(\bb^{\top}\bX\left(t\right));t\in[\tau_{0},\tau],\bb\in\mathcal{B}\} $
is a $\mathbf{P}$-Donsker class and Glivenko-Cantelli class for $u=0,1$.
Therefore, for $u=0,1$, $\sup_{t\in\left[\tau_{0},\tau\right],\bb\in\mathcal{B}}\sqrt{n}\left\Vert S^{\left(u\right)}\left(t,\bb\right)-s^{\left(u\right)}\left(t,\bb\right)\right\Vert_{u+1}\to0$
in probability as $n\to\infty$. By the continuous mapping theorem, $\bar{\bX}(t;\hat{\bb})$ converges to $s^{\left(1\right)}(t,\bb_{0})/s^{\left(0\right)}(t,\bb_{0})$ uniformly for $t\in[\tau_{0},\tau]$. Then, the dominated convergence theorem yields the convergence of $\bar{\bX}(t;\hat{\bb})$ in $\hat{d}$ and completes the second part proof of Lemma~\ref{lem:xbarconsist}.

\section{Additional Simulation Results}\label{app:addsim}

We present additional simulation results in this section.
We mainly consider generating $X_{1}(t)$ with various number of pieces while keeping other settings the same to those Section~\ref{sec:sim}.
In other words, we generate $X_{1}\left(t\right)=\sum_{j=1}^{M}X_{1j}I(U_{(j-1)}<t\le U_{(j)})$ where $\{X_{1j}\}_{j=1}^{M}\sim_{i.i.d.}\mathrm{Unif}\left(0,1\right)$, $\{U_{(j)}\}_{j=1}^{M}$ as an ordered i.i.d. sample from $\mathrm{Unif}\left(0,\tau\right)$ , $U_{(0)}=0$ and $U_{(M)}=\tau$.
Here, $M$ represents the number of pieces of  $X_{1}(t)$.
The simulation settings with larger $M$ should deviate from Condition~\ref{enu:Xconst} more.
In Tables~\ref{tab:simres_i4} and \ref{tab:simres_i15}, we present the results of the case where $X_{1}(t)$ was generated with $M=4$ and $M=15$ pieces, respectively.

In Table~\ref{tab:simres_time}, we present the mean of the respective ratios of the computational times of Robust and $\bar{\bX}$-based Robust to Profile. We consider number of covariates $p=1,2$ and $3$:
\begin{enumerate}
	\item $p=1$, $X_{1}(t)$ was generated with $M=10$ and $\beta_{1}=1$;
	\item $p=2$, $X_{1}(t)$ was generated with $M=10$, $X_2 \sim \mathrm{Unif}\left(0,1\right) $  and $\bb=(1,-1)^{\top}$;
	\item $p=3$, $X_{1}(t)$ was generated with $M=10$ and $X_2, X_3 \sim \mathrm{Unif}\left(0,1\right) $  with $\bb=(1,-1,0.5)^{\top}$.
\end{enumerate}
The computational times are evaluated by Rcpp \citep{Eddelbuettel2013} implementation running on Intel(R) Xeon(R) CPU E5-2687W v3.

\begin{table}[H]
	\centering{}\caption{Simulation results on the estimation of regression parameters with
		$\lambda\left(t\right)=4\left(\sin\left(4\pi t\right)+2\right)$ or $\lambda\left(t\right)=4\left(\sin\left(4\pi t\right)+2\right)$ when  $X_{1}(t)$ consists of 4 pieces.
		\label{tab:simres_i4}}
	
	\centering{}%
	\begin{tabular}{cccrrrrrrrr}
		\toprule 
		& Poisson  &  &  &  & \multicolumn{2}{c}{Robust ($\bar{\bX}$-based)} & \multicolumn{2}{c}{Info ($\bar{\bX}$-based)} & \multicolumn{2}{c}{Profile}\tabularnewline
		\cmidrule(lr){6-7} \cmidrule(lr){8-9} \cmidrule(lr){10-11}
		$n$  & Model  &  & \mc{Bias}  & \mc{SE}  & \mc{SEE}  & \mc{CP}  & \mc{SEE}  & \mc{CP}  & \mc{SEE}  & \mc{CP}\tabularnewline
		\midrule 
		\multicolumn{11}{c}{$\ensuremath{\lambda\left(t\right)=16/\left(1+t\right)}$}\tabularnewline
		100  & Yes  & $\hat{\beta}_{1}$  & <0.001 & 0.092 & 0.088(0.088) & 92.9(92.8) & 0.094(0.094) & 94.9(94.9) & 0.094 & 94.5\tabularnewline
		&  & $\hat{\beta}_{2}$  & 0.001 & 0.082 & 0.082(0.082) & 94.9(94.9) & 0.087(0.087) & 96.1(96.1) & 0.087 & 95.9\tabularnewline
		& No  & $\hat{\beta}_{1}$  & -0.037 & 0.315 & 0.300(0.299) & 93.3(93.1) & 0.031(0.031) & 15.7(15.7) & 0.031 & 15.7\tabularnewline
		&  & $\hat{\beta}_{2}$  & -0.008 & 0.376 & 0.371(0.371) & 93.7(93.6) & 0.021(0.021) & 7.8(7.8) & 0.021 & 7.8\tabularnewline
		\addlinespace
		200  & Yes  & $\hat{\beta}_{1}$  & -0.001 & 0.065 & 0.063(0.063) & 94.1(94.1) & 0.065(0.065) & 95.3(95.2) & 0.065 & 95.2\tabularnewline
		&  & $\hat{\beta}_{2}$  & <0.001 & 0.059 & 0.058(0.058) & 94.1(94.1) & 0.060(0.060) & 94.7(94.7) & 0.06 & 94.5\tabularnewline
		& No  & $\hat{\beta}_{1}$  & -0.017 & 0.221 & 0.211(0.211) & 92.5(92.5) & 0.021(0.021) & 14.8(14.8) & 0.021 & 14.9\tabularnewline
		&  & $\hat{\beta}_{2}$  & -0.003 & 0.271 & 0.263(0.263) & 92.5(92.5) & 0.014(0.014) & 8.4(8.4) & 0.014 & 8.4\tabularnewline
		\addlinespace
		400  & Yes  & $\hat{\beta}_{1}$  & 0.002 & 0.045 & 0.045(0.045) & 94.0(94.0) & 0.045(0.045) & 94.4(94.4) & 0.045 & 94.5\tabularnewline
		&  & $\hat{\beta}_{2}$  & 0.001 & 0.041 & 0.041(0.041) & 94.7(94.7) & 0.042(0.042) & 94.8(94.8) & 0.042 & 94.8\tabularnewline
		& No  & $\hat{\beta}_{1}$  & -0.009 & 0.15 & 0.150(0.150) & 93.6(93.5) & 0.014(0.014) & 15.8(15.8) & 0.014 & 15.7\tabularnewline
		&  & $\hat{\beta}_{2}$  & -0.002 & 0.183 & 0.187(0.187) & 94.1(94.1) & 0.010(0.010) & 7.9(7.9) & 0.01 & 7.9\tabularnewline
		\addlinespace
		\multicolumn{11}{c}{$\lambda\left(t\right)=4\left(\sin\left(4\pi t\right)+2\right)$}\tabularnewline
		100  & Yes  & $\hat{\beta}_{1}$  & -0.001 & 0.096 & 0.093(0.093) & 93.7(93.7) & 0.099(0.099) & 94.7(94.7) & 0.098 & 94.7\tabularnewline
		&  & $\hat{\beta}_{2}$  & 0.003 & 0.087 & 0.086(0.086) & 93.7(93.7) & 0.092(0.092) & 95.5(95.5) & 0.091 & 95.3\tabularnewline
		& No  & $\hat{\beta}_{1}$  & 0.003 & 0.299 & 0.291(0.290) & 93.3(93.3) & 0.034(0.034) & 18.7(18.7) & 0.034 & 18.5\tabularnewline
		&  & $\hat{\beta}_{2}$  & 0.010 & 0.384 & 0.368(0.367) & 92.2(92.2) & 0.023(0.023) & 8.6(8.6) & 0.023 & 8.6\tabularnewline
		200  & Yes  & $\hat{\beta}_{1}$  & 0.001 & 0.068 & 0.066(0.066) & 94.4(94.4) & 0.068(0.068) & 94.7(94.7) & 0.068 & 94.8\tabularnewline
		&  & $\hat{\beta}_{2}$  & 0.003 & 0.062 & 0.061(0.061) & 94.0(94.0) & 0.063(0.063) & 95.4(95.4) & 0.063 & 95.5\tabularnewline
		& No  & $\hat{\beta}_{1}$  & -0.001 & 0.207 & 0.209(0.209) & 94.9(94.9) & 0.023(0.023) & 15.6(15.5) & 0.023 & 15.6\tabularnewline
		&  & $\hat{\beta}_{2}$  & 0.008 & 0.267 & 0.263(0.263) & 92.7(92.7) & 0.015(0.015) & 8.1(8.1) & 0.015 & 8.2\tabularnewline
		400  & Yes  & $\hat{\beta}_{1}$  & <0.001 & 0.047 & 0.047(0.047) & 95.4(95.4) & 0.048(0.048) & 95.6(95.7) & 0.048 & 95.6\tabularnewline
		&  & $\hat{\beta}_{2}$  & 0.001 & 0.044 & 0.043(0.043) & 95.1(95.1) & 0.044(0.044) & 95.5(95.5) & 0.044 & 95.5\tabularnewline
		& No  & $\hat{\beta}_{1}$  & <0.001 & 0.145 & 0.147(0.147) & 95.0(95.0) & 0.016(0.016) & 14.4(14.4) & 0.016 & 14.4\tabularnewline
		&  & $\hat{\beta}_{2}$  & <0.001 & 0.189 & 0.186(0.186) & 94.5(94.5) & 0.010(0.010) & 7.3(7.3) & 0.011 & 7.3\tabularnewline
		\bottomrule
	\end{tabular}
\end{table}

\begin{table}[H]
	\centering{}\caption{Simulation results on the estimation of regression parameters with
		$\lambda\left(t\right)=4\left(\sin\left(4\pi t\right)+2\right)$ or
		$\lambda\left(t\right)=4\left(\sin\left(4\pi t\right)+2\right)$ when
		$X_{1}(t)$ consists of 15 pieces. \label{tab:simres_i15}}
	
	\centering{}%
	\begin{tabular}{cccrrrrrrrr}
		\toprule 
		& Poisson  &  &  &  & \multicolumn{2}{c}{Robust ($\bar{\bX}$-based)} & \multicolumn{2}{c}{Info ($\bar{\bX}$-based)} & \multicolumn{2}{c}{Profile}\tabularnewline
		\cmidrule(lr){6-7} \cmidrule(lr){8-9} \cmidrule(lr){10-11}
		$n$  & Model  &  & \mc{Bias}  & \mc{SE}  & \mc{SEE}  & \mc{CP}  & \mc{SEE}  & \mc{CP}  & \mc{SEE}  & \mc{CP}\tabularnewline
		\midrule 
		\multicolumn{11}{c}{$\ensuremath{\lambda\left(t\right)=16/\left(1+t\right)}$}\tabularnewline
		100  & Yes  & $\hat{\beta}_{1}$  & -0.014 & 0.116 & 0.106(0.105) & 91.7(91.7) & 0.114(0.114) & 93.6(93.6) & 0.113 & 93.2\tabularnewline
		&  & $\hat{\beta}_{2}$  & 0.002 & 0.085 & 0.082(0.082) & 93.0(93.0) & 0.087(0.087) & 94.5(94.5) & 0.086 & 94.3\tabularnewline
		& No  & $\hat{\beta}_{1}$  & -0.138 & 0.304 & 0.261(0.256) & 89.6(88.9) & 0.052(0.052) & 25.4(25.4) & 0.051 & 24.6\tabularnewline
		&  & $\hat{\beta}_{2}$  & 0.008 & 0.386 & 0.366(0.366) & 92.1(92.1) & 0.021(0.021) & 7.0(7.0) & 0.021 & 7.1\tabularnewline
		\addlinespace
		200  & Yes  & $\hat{\beta}_{1}$  & -0.008 & 0.082 & 0.076(0.076) & 93.1(93.1) & 0.079(0.079) & 93.8(93.7) & 0.079 & 93.8\tabularnewline
		&  & $\hat{\beta}_{2}$  & 0.001 & 0.06 & 0.058(0.058) & 94.0(94.0) & 0.060(0.060) & 94.9(94.9) & 0.06 & 94.9\tabularnewline
		& No  & $\hat{\beta}_{1}$  & -0.076 & 0.205 & 0.187(0.185) & 90.5(90.0) & 0.034(0.034) & 26.0(25.9) & 0.034 & 25.6\tabularnewline
		&  & $\hat{\beta}_{2}$  & 0.009 & 0.275 & 0.260(0.260) & 93.6(93.6) & 0.014(0.014) & 7.6(7.6) & 0.014 & 7.6\tabularnewline
		\addlinespace
		400  & Yes  & $\hat{\beta}_{1}$  & -0.004 & 0.057 & 0.054(0.054) & 93.6(93.6) & 0.056(0.056) & 94.1(94.1) & 0.056 & 94.2\tabularnewline
		&  & $\hat{\beta}_{2}$  & 0.001 & 0.041 & 0.041(0.041) & 94.8(94.8) & 0.042(0.042) & 95.6(95.6) & 0.042 & 95.6\tabularnewline
		& No  & $\hat{\beta}_{1}$  & -0.04 & 0.14 & 0.134(0.133) & 92.5(92.1) & 0.023(0.023) & 25.8(25.8) & 0.023 & 25.8\tabularnewline
		&  & $\hat{\beta}_{2}$  & 0.007 & 0.182 & 0.184(0.184) & 95.0(95.0) & 0.010(0.010) & 8.3(8.3) & 0.01 & 8.3\tabularnewline
		\addlinespace
		\multicolumn{11}{c}{$\lambda\left(t\right)=4\left(\sin\left(4\pi t\right)+2\right)$}\tabularnewline
		100  & Yes  & $\hat{\beta}_{1}$  & -0.018 & 0.118 & 0.111(0.110) & 92.1(92.0) & 0.119(0.119) & 93.8(93.8) & 0.118 & 93.2\tabularnewline
		&  & $\hat{\beta}_{2}$  & -0.001 & 0.088 & 0.086(0.086) & 94.6(94.6) & 0.092(0.092) & 96.6(96.6) & 0.091 & 96.8\tabularnewline
		& No  & $\hat{\beta}_{1}$  & -0.133 & 0.296 & 0.257(0.252) & 87.6(87.1) & 0.057(0.057) & 27.6(27.6) & 0.056 & 26.7\tabularnewline
		&  & $\hat{\beta}_{2}$  & -0.011 & 0.369 & 0.356(0.356) & 91.8(91.8) & 0.024(0.024) & 9.5(9.5) & 0.024 & 9.6\tabularnewline
		200  & Yes  & $\hat{\beta}_{1}$  & -0.007 & 0.084 & 0.079(0.079) & 93.8(93.8) & 0.083(0.083) & 95.0(95.0) & 0.082 & 94.7\tabularnewline
		&  & $\hat{\beta}_{2}$  & -0.001 & 0.063 & 0.061(0.061) & 94.1(94.1) & 0.063(0.063) & 95.1(95.1) & 0.063 & 94.9\tabularnewline
		& No  & $\hat{\beta}_{1}$  & -0.067 & 0.202 & 0.184(0.182) & 91.1(90.8) & 0.038(0.038) & 25.8(25.7) & 0.038 & 25.8\tabularnewline
		&  & $\hat{\beta}_{2}$  & -0.007 & 0.26 & 0.258(0.258) & 93.9(93.9) & 0.015(0.015) & 9.8(9.8) & 0.016 & 9.8\tabularnewline
		400  & Yes  & $\hat{\beta}_{1}$  & -0.003 & 0.061 & 0.057(0.057) & 92.5(92.5) & 0.058(0.058) & 93.5(93.5) & 0.058 & 93.4\tabularnewline
		&  & $\hat{\beta}_{2}$  & -0.003 & 0.043 & 0.043(0.043) & 95.0(95.0) & 0.044(0.044) & 95.4(95.4) & 0.044 & 95.3\tabularnewline
		& No  & $\hat{\beta}_{1}$  & -0.031 & 0.14 & 0.131(0.130) & 92.4(92.2) & 0.026(0.026) & 28.6(28.6) & 0.026 & 28.7\tabularnewline
		&  & $\hat{\beta}_{2}$  & -0.005 & 0.185 & 0.184(0.184) & 94.4(94.4) & 0.011(0.011) & 9.3(9.3) & 0.011 & 9.3\tabularnewline
		\bottomrule
	\end{tabular}
\end{table}

\begin{table}[H]
	\centering{}\caption{Simulation results on computational time comparison of variance estimators with
		$\lambda\left(t\right)=4\left(\sin\left(4\pi t\right)+2\right)$ or
		$\lambda\left(t\right)=4\left(\sin\left(4\pi t\right)+2\right)$ when
		$X_{1}(t)$ consists of 10 pieces. \label{tab:simres_time}}
	
	\centering{}%
	\begin{tabular}{crrrrrrrr}
		\toprule 
		Poisson &  & \multicolumn{3}{c}{Robust to Profile} &  & \multicolumn{3}{c}{$\bar{\bX}$-based Robust to Profile}\tabularnewline
		Model & $n$ & \multicolumn{1}{r}{$p=1$} & \multicolumn{1}{r}{$p=2$} & \multicolumn{1}{r}{$p=3$} &  & \multicolumn{1}{r}{$p=1$} & \multicolumn{1}{r}{$p=2$} & \multicolumn{1}{r}{$p=3$}\tabularnewline
		\midrule 
		& \multicolumn{8}{c}{$\ensuremath{\lambda\left(t\right)=16/\left(1+t\right)}$}\tabularnewline
		Yes & 100 & 3.67 & 3.82 & 3.42 &  & 82.94 & 110.81 & 120.52\tabularnewline
		& 200 & 3.61 & 3.76 & 3.29 &  & 39.62 & 39.57 & 40.77\tabularnewline
		& 400 & 3.25 & 3.41 & 3.05 &  & 15.82 & 22.91 & 24.18\tabularnewline
		& 800 & 2.60 & 2.80 & 2.53 &  & 9.22 & 11.79 & 12.85\tabularnewline
		& 1600 & 1.83 & 2.04 & 1.87 &  & 5.11 & 6.81 & 7.29\tabularnewline
		\addlinespace
		No & 100 & 4.42 & 4.33 & 4.19 &  & 100.95 & 127.34 & 150.25\tabularnewline
		& 200 & 4.57 & 4.45 & 3.97 &  & 49.94 & 45.33 & 49.68\tabularnewline
		& 400 & 4.30 & 3.93 & 3.62 &  & 19.79 & 26.15 & 29.27\tabularnewline
		& 800 & 3.53 & 3.20 & 3.07 &  & 12.45 & 13.60 & 15.49\tabularnewline
		& 1600 & 2.51 & 2.31 & 2.24 &  & 7.03 & 7.81 & 8.62\tabularnewline
		\addlinespace
		& \multicolumn{8}{c}{$\lambda\left(t\right)=4\left(\sin\left(4\pi t\right)+2\right)$}\tabularnewline
		Yes & 100  & 3.38 & 3.8 & 3.23 & & 75.2 & 97.52 & 98.96\tabularnewline
		& 200  & 3.26 & 3.64 & 3.10 & &37.29 & 36.29 & 35.54\tabularnewline
		& 400  & 3.10 & 3.24 & 2.84 & &14.25 & 21.71 & 20.96\tabularnewline
		& 800  & 2.41 & 2.65 & 2.39 & &8.89 & 10.63 & 11.74\tabularnewline
		& 1600  & 1.72 & 1.95 & 1.80 & &4.91 & 6.47 & 7.08\tabularnewline
		\addlinespace
		No & 100 & 3.78 & 4.24 & 4.07 & &90.15 & 115.78 & 123.15\tabularnewline
		& 200  & 4.08 & 4.19 & 3.74 & &46.40 & 41.78 & 43.10\tabularnewline
		& 400  & 3.82 & 3.7 & 3.36 & &18.63 & 24.54 & 25.07\tabularnewline
		& 800  & 3.13 & 2.96 & 2.84 & &11.33 & 12.00 & 14.10\tabularnewline
		& 1600  & 2.28 & 2.21 & 2.14 & &6.55 & 7.34 & 8.27\tabularnewline
		\bottomrule
	\end{tabular}
\end{table}
\newpage

\section{The estimates of the three baseline cumulative rate functions corresponding to three types of STIs}\label{app:reallam}

\begin{figure}[H]
	\centering{}\includegraphics[width=0.8\textwidth]{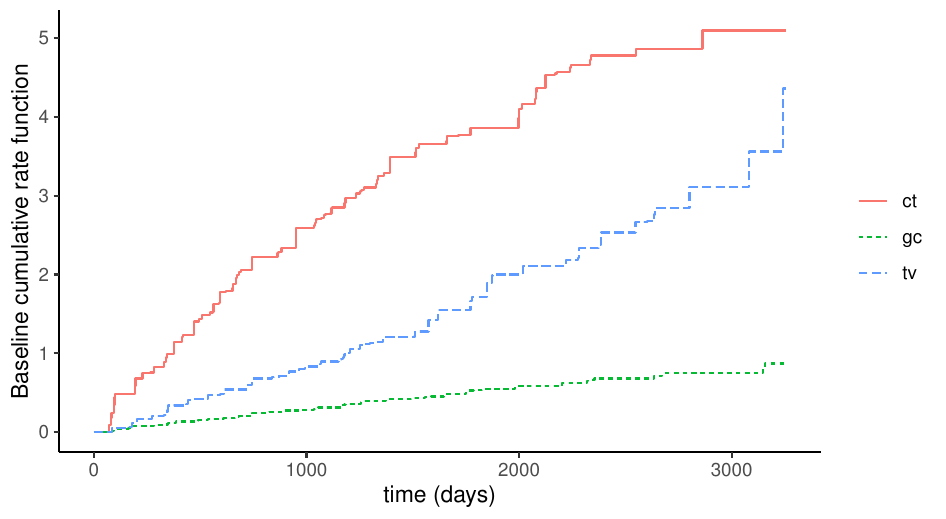}\caption{Estimates of the three baseline cumulative rate functions corresponding to three types of STIs.\label{fig:realdataLambdaplot} }
\end{figure}

To further investigate the three types of STIs, we obtained the estimated cumulative baseline rate functions 
corresponding to each and plotted them in Figure \ref{fig:realdataLambdaplot}.
They suggest that the CT infection 
was much more prevalent among young women than the NG and TV infections. 
The finding is highly consistent with the national data that indicate CT as the most prevalent STI among US women aged 14-39 years \citep{Torrone2014}. 
The estimated baseline mean functions under model \eqref{eq:meanmodel} all violated monotonicity and hence are not shown here.

\newpage

\section{The Detailed Proofs}\label{app:proofs}

\subsection{The final step of the proof of Theorem~\ref{thm:consistency}}

Since $\hat{\bL}\left(t\right)$ is non-decreasing on $[0,\tau]$, $\hat{\bL}\left(t\right)$
is also of bounded variation on $[0,\tau]$.
Since $\hat{\bL}\left(t\right)$ is uniformly bounded
a.s. $t\in\left[0,\tau\right]$ and $\Theta=\mathcal{B}\times\mathcal{F}$ is compact, the Helly selection theorem implies that $\hat{\bt}$ has a subsequence
of $\hat{\bt}_{n_{l}}$ that converges
to $\bt^{+}=\left(\bb^{+},\bL^{+}\right)$ as $l\to\infty$,
where $\bL^{+}$ is a non-decreasing bounded function defined
on $\text{\ensuremath{\left[0,\tau\right]}}$. Under \ref{enu:m0upperbound},
$\mathbf{P}m_{\bt}\left(O\right)<\infty$ for any $\bt\in\mathcal{B}\times\mathcal{F}$
because $m_{\bt}\left(O\right)\le M_{0}\left(O\right)$ by Jensen' inequality and $\mathbf{P}M_{0}<\infty$.
Besides, $m_{\bt}$ is upper semicontinuous in $\mathcal{F}$ for
each $O$ and $\mathcal{B}\times\mathcal{F}$ is compact. Therefore,
by the one-sided Gilvenko-Cantelli theorem \citep[Theorem A.1]{Wellner2000},
$
\lim\sup_{n\to\infty}\sup_{\bt\in\mathcal{B}\times\mathcal{F}}\left(\mathbb{P}-\mathbf{P}\right)m_{\bt}\le0
$
a.s.. This inequality implies, with probability 1, 
$
\lim\sup_{l\to\infty}\mathbb{M}_{n}(\hat{\bt}_{n_{l}})\le\lim\sup_{l\to\infty}\mathbf{M}(\hat{\bt}_{n_{l}})=\mathbf{M}(\bt^{+})
$
by the continuity of $\mathbf{M}$ in $\bt$ and the dominated
convergence theorem. By the law of large numbers, $\mathbb{M}_{n}\left(\bt_{0}\right)\to\mathbf{M}\left(\bt_{0}\right)$
almost surely.
By the fact that $\mathbb{M}_{n}(\bt_{0})\le\mathbb{M}_{n}(\hat{\bt})$,
we have, with probability 1, 
$
\mathbf{M}(\bt_{0}) \le\lim\inf_{n\to\infty}\mathbb{M}_{n}(\hat{\bt}) \le\lim\inf_{l\to\infty}\mathbb{M}_{n_{l}}(\hat{\bt}_{n_{l}}) \le\lim\sup_{l\to\infty}\mathbb{M}_{n}(\hat{\bt}_{n_{l}}) \le \mathbf{M}(\bt^{+})
$.
Thus, $\mathbf{M}\left(\bt_{0}\right)\le\mathbf{M}\left(\bt^{+}\right)$.
On the other hand, $\mathbf{M}\left(\bt_{0}\right)\ge\mathbf{M}\left(\bt^{+}\right)$ since $\bt_{0}$ is the unique maximizer of $\mathbf{M}(\cdot)$.
In then follows that $\mathbf{M}\left(\bt_{0}\right)=\mathbf{M}\left(\bt^{+}\right)$.
Recalling that $\bt_{0}$ is the unique maximizer of $\mathbf{M}(\cdot)$, then $\bb_{0}=\bb^{+}$
and $\bL_{0}=\bL^{+}$ a.e. on $\mu_{2}$. Since this is true
for any convergent subsequence of $\hat{\bt}$, we conclude that
all the limits of the subsequences of $\hat{\bt}_{n}$ are
equal to $\bt_{0}$.
In other words, we obtain the strong consistency of $\hat{\beta}$ and the almost everywhere convergence
of $\hat{\bL}\left(t\right)$ to $\bL_{0}\left(t\right)$
on $\mu_{2}$ almost surely.
Due to the uniform boundedness
of $\hat{\bL}$, the dominated convergence theorem yields the
strong consistency of $\hat{\bt}$ in the metric $d\left(\cdot,\cdot\right)$.

\subsection{Proof of Lemma~\ref{lem:bracketingnumber}}

Since $\bL^{\prime}$ is bounded and \ref{enu:mu2con}
holds, by the argument similar to \citet[p. 2127]{Wellner2007}, for any $\epsilon>0$,
there exists a set of brackets $\left[\Delta\bL_{1}^{l},\Delta\bL_{1}^{r}\right]$,
$\left[\Delta\bL_{2}^{l},\Delta\bL_{2}^{r}\right]$, ... $\left[\Delta\bL_{q_{1}}^{l},\Delta\bL_{q_{1}}^{r}\right]$
with $q_{1}\le\exp\left(M/\epsilon\right)$.
Here, for any $\bL\in\mathcal{F}_{\eta}$,
$\Delta\bL_{i}^{l}\left(u_{1},u_{2}\right)\le\Delta\bL\left(u_{1},u_{2}\right)\le\Delta\bL_{i}^{r}\left(u_{1},u_{2}\right)$
for all $u_{1},u_{2}\in\left[\tau_{0},\tau\right]$ and some $1\le k\le q_{1}$
with $\left\Vert \Delta\bL_{k}^{l}-\Delta\bL_{k}^{r}\right\Vert _{L_{2}\left(\mu_{1}\right)}\le\epsilon$.
In addition, since $\mathcal{B}$ is compact, we can construct
an $\epsilon$-net for $\mathcal{B}$, $\bb_{1},\ldots,\bb_{q_{2}}$
with $q_{2}\lesssim\left(1/\epsilon\right)^{p}$ such that for any
$\bb\in\mathcal{B}$, there exists $1\le l\le q_{2}$ such
that $\left|\bb^{\top}\bX\left(t\right)-\bb_{l}^{\top}\bX\left(t\right)\right|\le\epsilon$
and $|\exp(\bb^{\top}\bX(t))-\exp(\bb_{l}^{\top}\bX(t))|\le C_{2}\epsilon$.
Therefore, we can construct a set of brackets for $\mathcal{M}_{\eta}$
as $\left[m_{k,l}^{l}\left(O\right),m_{k,l}^{r}\left(O\right)\right]$
for $k=1,\ldots q_{1}$ and $l=1\ldots q_{2}$ where 
\begin{multline*}
	m_{k,l}^{l}\left(O\right)=\sum_{j=1}^{J}\Delta N_{\cdot j}\log\left\{ \int_{T_{\cdot\left(j-1\right)}}^{T_{\cdot j}}\exp\left(\bb_{l}^{\top}\bX\left(t\right)-C_{1}\epsilon\right)d\bL_{k}^{l}\left(t\right)\right\} \\ -\int_{T_{\cdot\left(j-1\right)}}^{T_{\cdot j}}\left\{ \exp\left(\bb_{l}^{\top}\bX\left(t\right)\right)+C_{2}\epsilon\right\} d\bL_{k}^{r}\left(t\right)-m_{\bt_{0}},
\end{multline*}
and 
\begin{multline*}
	m_{k,l}^{r}\left(O\right)=\sum_{j=1}^{J}\Delta N_{\cdot j}\log\left\{ \int_{T_{\cdot\left(j-1\right)}}^{T_{\cdot j}}\exp\left(\bb_{l}^{\top}\bX\left(t\right)+C_{1}\epsilon\right)d\bL_{k}^{r}\left(t\right)\right\} \\ -\int_{T_{\cdot\left(j-1\right)}}^{T_{\cdot j}}\left\{ \exp\left(\bb_{l}^{\top}\bX\left(t\right)\right)-C_{2}\epsilon\right\} d\bL_{k}^{l}\left(t\right)-m_{\bt_{0}}.
\end{multline*}
One challenge of the theoretical development of the panel count data
is that $f_{k,l}\left(O\right)=m_{k,l}^{r}\left(O\right)-m_{k,l}^{l}\left(O\right)$
is generally not bounded since $\Delta N_{\cdot j}$ is not bounded.
To circumvent this, we resort to the $\epsilon$-bracketing number $N_{[\,]}\left(\epsilon,\mathcal{M}_{\eta},\left\Vert \cdot\right\Vert _{P,B}\right)$
using the Bernstein norm $\left\Vert \cdot\right\Vert _{P,B}$ where
$\Vert f \Vert _{P,B}=\{ 2\mathbf{P}(\exp(|f|)-1-|f|)\} ^{1/2}\le\{ \mathbf{P}(\exp(|f|)|f|^{2})\} ^{1/2}$
\citep[page 324]{Vaart1996}.
It follows that
\begin{multline*}
	\left|f_{k,l}\left(O\right)\right|=  2C_{1}\epsilon N_{\cdot J}+N_{\cdot J}\sum_{j=1}^{J}\left|\log\left\{ \frac{\int_{T_{\cdot\left(j-1\right)}}^{T_{\cdot j}}\exp\left(\bb_{l}^{\top}\bX\left(t\right)\right)d\bL_{k}^{r}\left(t\right)}{\int_{T_{\cdot\left(j-1\right)}}^{T_{\cdot j}}\exp\left(\bb_{l}^{\top}\bX\left(t\right)\right)d\bL_{k}^{l}\left(t\right)}\right\} \right|\\
	+\sum_{j=1}^{J}\left|\int_{T_{\cdot\left(j-1\right)}}^{T_{\cdot j}}\exp\left(\bb_{l}^{\top}\bX\left(t\right)\right)d\left(\bL_{k}^{r}\left(t\right)-\bL_{k}^{l}\left(t\right)\right)\right|+C_{2}\epsilon\sum_{j=1}^{J}\int_{T_{\cdot\left(j-1\right)}}^{T_{\cdot j}}d\left(\bL_{k}^{r}\left(t\right)+\bL_{k}^{l}\left(t\right)\right).
\end{multline*}
By an argument similar to Section 5 of \citet[p. 2128]{Wellner2007},
we can derive 
\[
\log\left\{ \frac{\int_{T_{\cdot\left(j-1\right)}}^{T_{\cdot j}}\exp\left(\bb_{l}^{\top}\bX\left(t\right)\right)d\bL_{k}^{r}\left(t\right)}{\int_{T_{\cdot\left(j-1\right)}}^{T_{\cdot j}}\exp\left(\bb_{l}^{\top}\bX\left(t\right)\right)d\bL_{k}^{l}\left(t\right)}\right\} \le\gamma_{0}^{-1}\left(\int_{T_{\cdot\left(j-1\right)}}^{T_{\cdot j}}\exp\left(\bb_{l}^{\top}\bX\left(t\right)\right)d\left(\bL_{k}^{r}\left(t\right)-\bL_{k}^{l}\left(t\right)\right)\right),
\]
for some finite positive $\gamma_{0}$.
Therefore, under \ref{enu:paraspace}, \ref{enu:xbounded}, \ref{enu:separable} and \ref{enu:finite_obs},
\begin{align*}
	\left\Vert f_{k,l}\left(O\right)\right\Vert _{P,B}^{2}  \le & \mathbf{P}\exp\left(\left|f_{k,l}\left(O\right)\right|\right)\left|f_{k,l}\left(O\right)\right|^{2}\\
	\lesssim & \mathbf{P}\ensuremath{\exp\left(2C_{1}\epsilon N_{\cdot J}\right)\left\{ N_{\cdot J}^{2}\sum_{j=1}^{J}\log\left\{ \frac{\int_{T_{\cdot\left(j-1\right)}}^{T_{\cdot j}}\exp\left(\bb_{l}^{\top}\bX\left(t\right)\right)d\bL_{k}^{r}\left(t\right)}{\int_{T_{\cdot\left(j-1\right)}}^{T_{\cdot j}}\exp\left(\bb_{l}^{\top}\bX\left(t\right)\right)d\bL_{k}^{l}\left(t\right)}\right\} ^{2}\right.}\\
	&+\left.\left(\int_{T_{\cdot\left(j-1\right)}}^{T_{\cdot j}}\exp\left(\bb_{l}^{\top}\bX\left(t\right)\right)d\left(\bL_{k}^{r}\left(t\right)-\bL_{k}^{l}\left(t\right)\right)\right)^{2}+C_{3}\epsilon^{2}\right\} .
\end{align*}
By \ref{enu:xbounded} and \ref{enu:expbound}, we further have 
$
\left\Vert f_{k,l}\left(O\right)\right\Vert _{P,B}^{2}  \lesssim\mathbf{P}\sum_{j=1}^{J}\left(\int_{T_{\cdot\left(j-1\right)}}^{T_{\cdot j}}d\left(\bL_{k}^{r}\left(t\right)-\bL_{k}^{l}\left(t\right)\right)\right)^{2}+\epsilon^{2}=\left\Vert \Delta\bL_{k}^{l}-\Delta\bL_{k}^{r}\right\Vert _{L_{2}\left(\mu_{1}\right)}+ \epsilon^{2}.
$
It then follows that $\left\Vert f_{k,l}\left(O\right)\right\Vert _{P,B}^{2} \lesssim\epsilon^{2}$.
Thus, $\epsilon$-bracketing number for $\mathcal{M}_{\eta}$ will
be of the order $\left(1/\epsilon\right)^{d}\exp\left(C/\epsilon\right)$ for some constant $C>0$.
It follows easily that $\log N_{[\,]}\left(\epsilon,\mathcal{M}_{\eta},\left\Vert \cdot\right\Vert _{P,B}\right)\lesssim1/\epsilon$.

\subsection{Proof of Proposition~\ref{prop:genasym}}

\begin{proof}
	By conditions (a4) and (a5) with $S\left(\bt_{0}\right)=0$ from
	condition (a2), 
	\begin{align*}
		-\sqrt{n}\dot{S}(\bt_{0})(\hat{\bt}_{n}-\bt_{0})[h_{1},h_{2}] & =-\sqrt{n}(S(\hat{\bt}_{n})-S(\bt_{0}))[h_{1},h_{2}]+o_{p}(1)\\
		& =-\sqrt{n}S(\hat{\bt}_{n})[h_{1},h_{2}]-\sqrt{n}S(\bt_{0})[h_{1},h_{2}]+o_{p}(1)\\
		& =-\sqrt{n}S(\hat{\bt}_{n})[h_{1},h_{2}]+o_{p}(1).
	\end{align*}
	By conditions (a1) and $S_{n}(\hat{\bt}_{n})=o_{p}\left(n^{-1/2}\right)$
	in (a2), 
	\[
	-\sqrt{n}S(\hat{\bt}_{n})\left[h_{1},h_{2}\right]=\sqrt{n}\left(S_{n}-S\right)\left(\bt_{0}\right)\left[h_{1},h_{2}\right]+o_{p}\left(1\right).
	\]
	Combining the last two equations, the proposition follows.
\end{proof}

\section{Detailed derivation of $\boldsymbol{\Omega}$}\label{app:omega}

By the definition of $\bS$ in terms of the EM algorithm,
\begin{align*}
	-\bO= & E\left[\sum_{j_{1}=1}^{J}\sum_{T_{\cdot\left(j_{1}-1\right)}<t_{k_{1}}\le T_{\cdot j_{1}}}\sum_{j_{1}=1}^{J}\sum_{T_{\cdot\left(j_{1}-1\right)}<t_{k_{2}}\le T_{\cdot j_{1}}}\right.\\
	& \times E\left[\left(\hat{E}\left[W_{\cdot k_{1}}|\bb_{0},\bl_{0}\right]-\Delta_{\cdot k_{1}}\exp\left(\bb_{0}^{\top}\bX\left(t_{k_{1}}\right)\right)\lambda_{0k_{1}}\right)\right.\\
	& \times\left.\left(\hat{E}\left[W_{\cdot k_{2}}|\bb_{0},\bl_{0}\right]-\Delta_{\cdot k_{1}}\exp\left(\bb_{0}^{\top}\bX\left(t_{k_{2}}\right)\right)\lambda_{0k_{2}}\right)|\bX,\Delta_{\cdot k_{1}},\Delta_{\cdot k_{2}}\right]\\
	& \times\left.\left(\bX\left(t_{k_{1}}\right)-\bar{\bx}\left(t_{k}\right)\right)\left(\bX\left(t_{k_{1}}\right)-\bar{\bx}\left(t_{k}\right)\right)\right],
\end{align*}
where 
\[
\hat{E}\left[W_{\cdot k}|\bb_{0},\bl_{0}\right]=\frac{\Delta_{\cdot k}\exp\left(\bb_{0}^{\top}\bX\left(t_{k}\right)\right)\lambda_{0k}}{\sum_{T_{\cdot\left(j-1\right)}<t_{l}\le T_{\cdot j}}\Delta_{\cdot l}\exp\left(\bb_{0}^{\top}\bX\left(t_{l}\right)\right)\lambda_{0l}}\Delta N_{\cdot j},
\]
for $T_{\cdot\left(j-1\right)}<t_{k}\le T_{\cdot j}$. Under model
\eqref{eq:ratemodel}, 
$
E\left[\Delta N_{\cdot j}|\bX,\Delta\right]=\sum_{T_{\cdot\left(j-1\right)}<t_{l}\le T_{\cdot j}}\Delta_{\cdot l}\exp\left(\bb_{0}^{\top}\bX\left(t_{l}\right)\right)\lambda_{0l}
$
and 
$
E\left[\hat{E}\left[W_{\cdot k}|\bb_{0},\bl_{0}\right]|\bX,\Delta_{\cdot k}\right]=\Delta_{\cdot k}\exp\left(\bb_{0}^{\top}\bX\left(t_{k}\right)\right)\lambda_{0k}$.
Hence,
\begin{align*}
	& E\left[\left(\hat{E}\left[W_{\cdot k_{1}}|\bb_{0},\bl_{0}\right]-\Delta_{\cdot k_{1}}\exp\left(\bb_{0}^{\top}\bX\left(t_{k_{1}}\right)\right)\lambda_{0k_{1}}\right)\right.\\
	& \times\left.\left(\hat{E}\left[W_{\cdot k_{2}}|\bb_{0},\bl_{0}\right]-\Delta_{\cdot k_{1}}\exp\left(\bb_{0}^{\top}\bX\left(t_{k_{2}}\right)\right)\lambda_{0k_{2}}\right)|\bX,\Delta_{\cdot k_{1}},\Delta_{\cdot k_{2}}\right]\\
	= & \mathrm{Cov}\left(\hat{E}\left[W_{\cdot k_{1}}|\bb_{0},\bl_{0}\right],\hat{E}\left[W_{\cdot k_{2}}|\bb_{0},\bl_{0}\right]|\bX\right)\\
	= & \mathrm{Cov}\left(\Delta N_{\cdot j_{1}},\Delta N_{\cdot j_{2}}|\bX,\Delta_{\cdot k_{1}},\Delta_{\cdot k_{2}}\right)\frac{\Delta_{\cdot k_{1}}\exp\left(\bb_{0}^{\top}\bX\left(t_{k_{1}}\right)\right)\lambda_{0k_{1}}}{\sum_{T_{\cdot\left(j_{1}-1\right)}<t_{l}\le T_{\cdot j_{1}}}\Delta_{\cdot l}\exp\left(\bb_{0}^{\top}\bX\left(t_{l}\right)\right)\lambda_{0l}}\\
	& \times\frac{\Delta_{\cdot k_{2}}\exp\left(\bb_{0}^{\top}\bX\left(t_{k_{2}}\right)\right)\lambda_{0k_{2}}}{\sum_{T_{\cdot\left(j_{2}-1\right)}<t_{l}\le T_{\cdot j_{2}}}\Delta_{\cdot l}\exp\left(\bb_{0}^{\top}\bX\left(t_{l}\right)\right)\lambda_{0l}}.
\end{align*}
If the Poisson model holds, we further have 

\[
\mathrm{Cov}\left(\Delta N_{\cdot j_{1}},\Delta N_{\cdot j_{2}}|\bX,\Delta_{\cdot k_{1}},\Delta_{\cdot k_{2}}\right)=\begin{cases}
	\sum_{T_{\cdot\left(j-1\right)}<t_{l}\le T_{\cdot j}}\Delta_{\cdot l}\exp\left(\bb_{0}^{\top}\bX\left(t_{l}\right)\right)\lambda_{0l} & \text{if }j_{1}=j_{2}=j;\\
	0 & \text{if }j_{1}\ne j_{2}.
\end{cases}
\]
because the counts of Poisson process occurrences in non-overlapping observation
intervals are mutually independent. Then, all the terms with $j_{1}\ne j_{2}$ in $\bS$ vanish.
Consequently, 
\begin{align*}
	-\bO & =E\left[\sum_{j=1}^{J}\sum_{T_{\cdot\left(j-1\right)}<t_{k}\le T_{\cdot j}}\frac{\left(\Delta_{\cdot k}\exp\left(\bb_{0}^{\top}\bX\left(t_{k}\right)\right)\lambda_{0k}\right)^{2}}{\sum_{T_{\cdot\left(j-1\right)}<t_{l}\le T_{\cdot j}}\Delta_{\cdot l}\exp\left(\bb_{0}^{\top}\bX\left(t_{l}\right)\right)\lambda_{0l}}\left(\bX\left(t_{k}\right)-\bar{\bx}\left(t_{k}\right)\right)^{\otimes2}\right].
\end{align*}

\bibliographystyle{plainnat}
\bibliography{ref}

\end{document}